\documentclass[reprint,superscriptaddress,amsmath,amssymb,aps,prb, longbibliography]{revtex4-1}

\usepackage{graphicx}
\usepackage{bm}
\usepackage{hyperref}
\usepackage{xr-hyper}
\usepackage{xcolor}
\usepackage{orcidlink}

\usepackage[normalem]{ulem}

\begin{document}

\title{Multiorbital Anisotropy and Magnetic Excitations in Trilayer Nickelate Pr$_4$Ni$_3$O$_{10}$}

\author{Ziqiang Guan\,\orcidlink{0000-0002-1097-6766}}
\thanks{these authors contributed equally}
\affiliation{Department of Physics, Harvard University, Cambridge, Massachusetts 02138, USA}

\author{Sophia F. R. TenHuisen\,\orcidlink{0000-0002-4379-7035}}
\thanks{these authors contributed equally}
\affiliation{Department of Physics, Harvard University, Cambridge, Massachusetts 02138, USA}
\affiliation{John A. Paulson School of Engineering and Applied Sciences, Harvard University, Cambridge, Massachusetts 02138, USA}

\author{Xuefei Guo\,\orcidlink{0000-0003-1088-6039}}
\affiliation{Condensed Matter Physics and Materials Science Department, Brookhaven National Laboratory, Upton, New York 11973, USA}

\author{Wei He\,\orcidlink{0000-0003-3522-3899}}
\affiliation{Condensed Matter Physics and Materials Science Department, Brookhaven National Laboratory, Upton, New York 11973, USA}
\affiliation{Stanford Institute for Materials and Energy Sciences,
SLAC National Accelerator Laboratory, Menlo Park, CA 94025, USA}

\author{Aulden K. Jones\,\orcidlink{0009-0000-4883-2357}}
\affiliation{Department of Physics, Harvard University, Cambridge, Massachusetts 02138, USA}

\author{Marli dos Reis Cantarino\,\orcidlink{0000-0002-6551-5481}}
\affiliation{European Synchrotron Radiation Facility, 71 Avenue des Martyrs, CS 40220, 38043 Grenoble Cedex 9, France}

\author{Xinglong Chen\,\orcidlink{0000-0003-2283-7293}}
\author{Hong Zheng\,\orcidlink{0000-0002-6342-7277}}
\author{J. F. Mitchell\,\orcidlink{0000-0002-8416-6424}}
\affiliation{Materials Science Division, Argonne National Laboratory, Lemont, Illinois 60439, USA}

\author{Nicholas B. Brookes\,\orcidlink{0000-0002-1342-9530}}
\affiliation{European Synchrotron Radiation Facility, 71 Avenue des Martyrs, CS 40220, 38043 Grenoble Cedex 9, France}

\author{Mark P. M. Dean\,\orcidlink{0000-0001-5139-3543}}
\affiliation{Condensed Matter Physics and Materials Science Department, Brookhaven National Laboratory, Upton, New York 11973, USA}

\author{Matteo Mitrano\,\orcidlink{0000-0002-0102-0391}}
\email{mmitrano@g.harvard.edu}
\affiliation{Department of Physics, Harvard University, Cambridge, Massachusetts 02138, USA}

\date{\today}

\begin{abstract}
Rare-earth substitution strongly modifies the structure, metallicity, and superconducting behavior of Ruddlesden--Popper (RP) nickelates, but its impact on their magnetic interactions remains unclear. Here, we combine x-ray absorption spectroscopy (XAS) and resonant inelastic x-ray scattering (RIXS) to resolve the multiorbital electronic structure and magnetic excitations of Pr$_4$Ni$_3$O$_{10}$. The absence of an overlapping rare-earth resonance provides direct access to the Ni $L_3$ edge, while polarization-dependent measurements at the Ni $L$- and O $K$-edges resolve distinct planar and out-of-plane Ni--O hybridization channels. Momentum-resolved RIXS reveals dispersive magnetic excitations that soften toward the incommensurate spin-density-wave wave vector. An effective spin-wave analysis yields a sizable intra-trilayer exchange coupling, comparable to the leading in-plane interactions and robust against rare-earth substitution. Our results identify the multilayer architecture, rather than rare-earth substitution alone, as a key control knob of the magnetic interactions in these materials.\vspace{-3mm}
\end{abstract}

\maketitle

\section{Introduction}\vspace{-3mm}

The emergence of high-temperature superconductivity in bilayer and trilayer Ruddlesden–Popper (RP) nickelates under pressure or epitaxial strain raises questions about how the multilayer Ni–O architecture controls electronic and magnetic interactions~\cite{sun2023signatures, zhu2024superconductivity, zhang2025bulk, wang2025recent, chen2024electronic}. Unlike cuprates or square-planar nickelates, these octahedrally coordinated compounds host both active Ni \(3d_{x^2-y^2}\) and \(3d_{3z^2-r^2}\) orbitals, with the latter coupling neighboring NiO$_2$ planes through apical oxygen~\cite{zhang2017large, puphal2026superconductivity}. 
Across the RP series, the number of layers changes the nominal Ni orbital filling and dimensionality of the electronic structure~\cite{greenblatt1997ruddlesden}. In bilayer La$_3$Ni$_2$O$_7$, the two NiO$_2$ planes form strongly coupled $3d_{3z^2-r^2}$ bonding and antibonding states, which play an important role in its low-energy electronic and magnetic response~\cite{luo2023bilayer,zhang2023electronic}.
Adding a third NiO$_2$ plane produces an additional nonbonding component and creates inequivalent inner and outer Ni sites, which in turn give rise to layer-dependent orbital occupation and magnetic response~\cite{zhang2020intertwined, samarakoon2023bootstrapped, labollita2024electronic, guan2026electronic}. 

Beyond the number and stacking sequence of the NiO$_2$ planes~\cite{zhu2024superconductivity, chen2024polymorphism, li2024design, shi2025pressure}, isovalent rare-earth substitution offers a complementary route for tuning the local crystal structure and electronic interactions through chemical pressure~\cite{wang2024bulk, wang2025chemical, li2026bulk}. This is particularly evident in the comparison between La$_4$Ni$_3$O$_{10}$ and Pr$_4$Ni$_3$O$_{10}$, where the smaller Pr ion contracts the lattice and distorts the Ni-O-Ni bond geometry~\cite{zhang2020high, huangfu2020anisotropic, rout2020structural}. Under pressure, bulk superconductivity in Pr$_4$Ni$_3$O$_{10}$ reaches an onset temperature near 40~K, surpassing the approximately 30~K onset reported for the La compound~\cite{zhang2025bulk, chen2025low, pei2026unveiling}. The enhanced superconducting transition temperature of Pr$_4$Ni$_3$O$_{10}$ is accompanied by a more robust density-wave state, which appears near 160~K at ambient pressure rather than at approximately 140~K in La$_4$Ni$_3$O$_{10}$ and persists over a broader pressure range~\cite{zhang2025bulk, chen2025low, khasanov2026multiple}. These contrasting trends underscore the sensitivity of trilayer nickelates to subtle structural changes and raise the question of how rare-earth substitution modifies their underlying electronic and magnetic interactions.

\begin{figure}[t]
    \centering
    \includegraphics[width=\columnwidth]{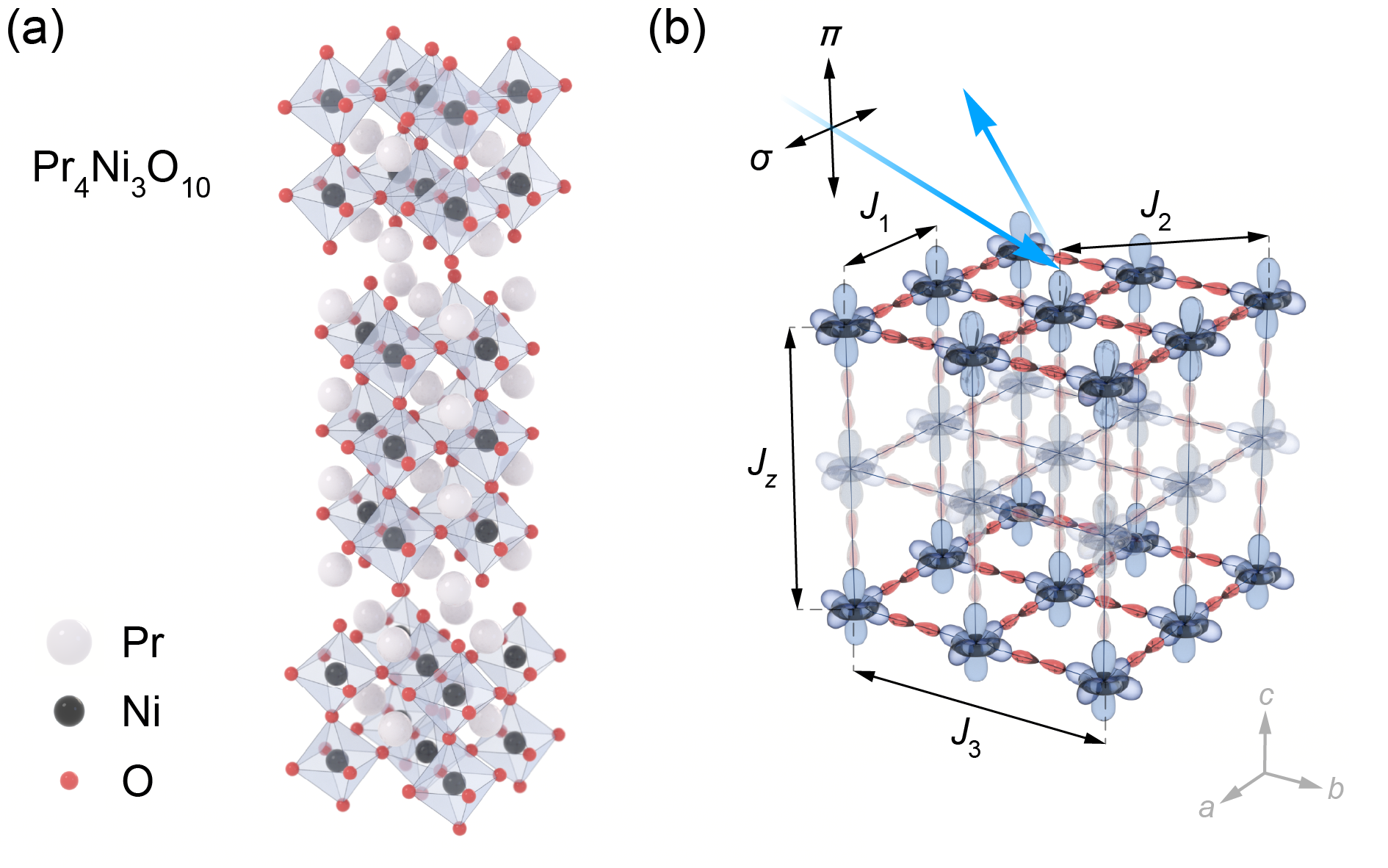}\vspace{-3mm}
    \caption{(Color online) Crystal structure and magnetic interactions in $\mathrm{Pr}_4\mathrm{Ni}_3\mathrm{O}_{10}$. (a) Crystal structure of $\mathrm{Pr}_4\mathrm{Ni}_3\mathrm{O}_{10}$, comprising trilayer blocks of corner-sharing $\mathrm{NiO}_6$ octahedra separated by Pr--O rock-salt layers. Pr, Ni, and O atoms are shown in light gray, black, and red, respectively. (b) Effective magnetic-interaction model and RIXS scattering geometry. The in-plane interactions $J_1$, $J_2$, and $J_3$ connect first-, second-, and third-neighbor Ni sites, respectively, within the outer layers carrying the ordered moment, while $J_z$ denotes the effective coupling between the two outer-layer SDW components. The transparent inner layer denotes suppressed magnetic moments. Blue arrows indicate the incident and scattered x-ray beams, with the $\sigma$- and $\pi$-polarization directions shown relative to the scattering plane. The orbital lobes represent the multiorbital Ni--O hybridization channels contributing to the in- and out-of-plane interactions.}
    \label{fig:structure_model}\vspace{-3mm}
\end{figure}

XAS and RIXS provide direct access to both the multiorbital electronic structure and magnetic interactions of RP nickelates. Ni $L$- and O $K$-edge measurements have identified substantial oxygen-ligand contributions and an active manifold of Ni $3d_{x^2-y^2}$, $3d_{3z^2-r^2}$, and O $2p$ states involving both in-plane and out-of-plane Ni--O hybridization~\cite{hepting2021soft,lin2021strong,shen2022role,chen2024electronic,zhang2026distinct,he2026layer}. Momentum-resolved RIXS has further revealed dispersive magnetic excitations and a strong evolution of magnetic interactions with layer number~\cite{lu2021magnetic,tenhuisen2025magnetic,chen2026dissecting,chan2026collective,he2026layer}. In bilayer La$_3$Ni$_2$O$_7$, the magnetic response is dominated by interlayer coupling through apical oxygens, whereas in trilayer La$_4$Ni$_3$O$_{10}$ the effective interlayer interaction is reduced to a magnitude comparable to the leading in-plane interactions. Whether this balance survives the structural and electronic changes introduced by Pr substitution remains unclear. Pr$_4$Ni$_3$O$_{10}$ also offers cleaner access to the trilayer Ni $L_3$-edge excitations because it avoids the La $M_4$-Ni $L_3$ overlap that complicates the interpretation of the Ni response in La-based RP nickelates~\cite{norman2023orbital,chen2024electronic}.

Here, we use XAS and RIXS to investigate the electronic and magnetic excitations of Pr$_4$Ni$_3$O$_{10}$. Polarization-dependent measurements at the Ni $L$- and O $K$-edges reveal clear anisotropy between in-plane and out-of-plane Ni--O hybridized states. Momentum-resolved Ni $L_3$-edge RIXS reveals dispersive low-energy magnetic excitations that soften toward the incommensurate SDW wave vector. Spin-wave modeling of their dispersion further indicates a sizable effective coupling between the two outer NiO$_2$ layers, comparable to the leading in-plane interactions. Together, these results reveal the multiorbital electronic response and magnetic interactions of Pr$_4$Ni$_3$O$_{10}$ and show that the dominant magnetic interactions within the trilayers persist despite the structural and electronic changes introduced by Pr substitution.

\vspace{-3mm}
\section{Experimental Methods}\vspace{-3mm}

High-quality single crystals of Pr$_4$Ni$_3$O$_{10}$ were synthesized by high-oxygen-pressure optical floating-zone growth following Ref.~\onlinecite{zhang2020high}. At ambient pressure, Pr$_4$Ni$_3$O$_{10}$ crystallizes in the monoclinic space group $P2_1/a$. Throughout this work, we use a pseudo-tetragonal notation with $a=b=3.829$~\AA{} and $c=27.5$~\AA{}, and express momentum transfer as $\mathbf{q}=(H,K,L)$ in reciprocal lattice units (r.l.u.).

We performed XAS and RIXS measurements at the ID32 beamline of the European Synchrotron Radiation Facility (ESRF) under proposal HC-6368~\cite{brookes2018beamline}. Polarization-dependent XAS and incident-energy-dependent RIXS were measured with $\sigma$- and $\pi$-polarized incident x-rays at $\theta=15^\circ$ in the $(H,H,L)$ scattering plane. Momentum-dependent Ni $L_3$-edge RIXS was measured with $\pi$-polarized incident x-rays in the $(H,H,L)$ and $(H,0,L)$ scattering planes. For the nominal in-plane scans, we fixed the spectrometer arm at $2\Theta=149.5^\circ$ and varied the sample angle to scan $q_{\parallel}=(H,H)$ and $(H,0)$, such that the out-of-plane $L$ component varied continuously along each trajectory. To probe the out-of-plane dependence, we varied $2\Theta$ and the sample angle together while fixing $q_{\parallel}=(-0.08,-0.08)$ r.l.u. and $(-0.10,0)$  r.l.u., respectively. The momentum-dependent measurements had an energy resolution of 24.5~meV. All experiments were performed at 30~K, above the $\sim26$~K transition coupling Pr and Ni magnetic moments~\cite{samarakoon2023bootstrapped}, and thus predominantly probe the Ni-sublattice density-wave state. We provide further experimental and analysis details in the Supplemental Material~\cite{supp}.

\vspace{-3mm}
\section{Multiorbital electronic structure}\vspace{-3mm}

Pr$_4$Ni$_3$O$_{10}$ consists of trilayer blocks of corner-sharing NiO$_6$ octahedra separated by Pr--O rock-salt layers [Fig.~\ref{fig:structure_model}(a)]. Relative to La$_4$Ni$_3$O$_{10}$, the Pr compound has a more compact but more distorted trilayer structure and considerably lower in-plane metallicity~\cite{zhang2020high,liu2026interlayer}. Below the density-wave transition near 160~K, intertwined charge- and spin-density-wave (SDW) order develops, with the charge modulation in phase across the trilayer and the antiphase SDW concentrated on the two outer layers, leaving the inner layer near a magnetic node~\cite{zhang2020intertwined,samarakoon2023bootstrapped,khasanov2026pressure}. Together with the intrinsic layer differentiation produced by $3d_{3z^2-r^2}$ hybridization, a density-wave-driven redistribution of $3d_{3z^2-r^2}$ occupation may further enhance this nonuniform layer response~\cite{guan2026electronic,labollita2024electronic,zhang2025spin}.

Figure~\ref{fig:structure_model}(b) summarizes the RIXS scattering geometry and orbital channels relevant to the polarization-dependent measurements. In our geometry, the electric field of $\sigma$-polarized light lies in the $ab$ plane and predominantly probes in-plane states, whereas $\pi$-polarized light has a dominant $c$-axis component and is more sensitive to out-of-plane states. Microscopically, Ni $3d_{x^2-y^2}$ states hybridize primarily with planar O $2p_{x,y}$ orbitals, while Ni $3d_{3z^2-r^2}$ states provide an out-of-plane channel through apical O $2p_z$ orbitals~\cite{chen2024electronic,zhang2026distinct,he2026layer}. The polarization dependence therefore directly distinguishes the planar and out-of-plane Ni--O hybridized states in Pr$_4$Ni$_3$O$_{10}$. Figure~\ref{fig:structure_model}(b) also shows the effective exchange interactions $J_1$, $J_2$, $J_3$, and $J_z$ discussed in the next section, with $J_z$ coupling the two outer layers across the magnetically suppressed inner layer.

\begin{figure}[t]
    \centering
    \includegraphics[width=\columnwidth]{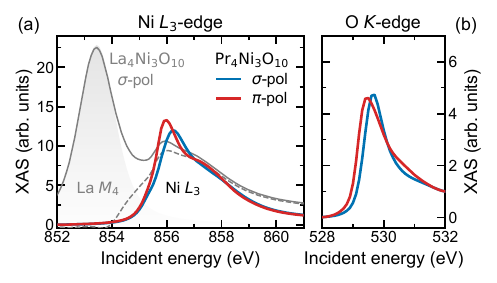}\vspace{-3mm}
    \caption{ (Color online) Polarization-dependent x-ray absorption spectra of Pr$_4$Ni$_3$O$_{10}$. (a) Ni $L_3$-edge XAS measured with $\sigma$- (blue) and $\pi$-polarized (red) incident x-rays. The corresponding $\sigma$-polarized spectrum of La$_4$Ni$_3$O$_{10}$ (solid gray line, from Ref.~\cite{he2026layer}) is shown for comparison. In the La compound, the La $M_4$ absorption overlaps the low-energy side of the Ni $L_3$ resonance, and the dashed gray curve shows the estimated Ni contribution after subtraction of the fitted La $M_4$ background. (b) Polarization-dependent O $K$-edge XAS of Pr$_4$Ni$_3$O$_{10}$. The energy shift and line-shape difference between $\sigma$- and $\pi$-polarized pre-edge spectra reveal anisotropy of the unoccupied oxygen-hybridized states.}\vspace{-3mm}
    \label{fig:xas}
\end{figure}

We first compare the Ni $L_3$-edge XAS of Pr$_4$Ni$_3$O$_{10}$ and La$_4$Ni$_3$O$_{10}$ [Fig.~\ref{fig:xas}(a)]. Both compounds share the same trilayer architecture and formal Ni valence of $+2.67$, corresponding to a nominal $3d^{7.33}$ configuration~\cite{zhang2020intertwined}. Strong Ni--O covalency, however, gives the ground state substantial ligand-hole character, as established by previous spectroscopic studies~\cite{chen2024electronic,tenhuisen2025magnetic,he2026layer}. In La$_4$Ni$_3$O$_{10}$, a strong La $M_4$-edge contribution overlaps the low-energy side of the Ni $L_3$-edge resonance and must be modeled and subtracted to isolate the Ni response~\cite{norman2023orbital,chen2026dissecting,chan2026collective}. After subtraction, the La and Pr compounds exhibit similar Ni $L_3$ line shapes, while Pr$_4$Ni$_3$O$_{10}$ provides direct access to the Ni resonance without overlapping rare-earth states. Because the Pr and La spectra were measured at different beamlines with different absolute energy calibrations, we compare only relative energies and line shapes. We provide details for the La $M_4$ subtraction and Pr--La comparison in Supplemental Material Sec.~S4.

Polarization-dependent XAS further reveals the multiorbital character of the unoccupied Ni--O states. At the O $K$ edge [Fig.~\ref{fig:xas}(b)], a pronounced pre-edge appears for both $\sigma$ and $\pi$ polarizations, reflecting ligand-hole states formed by strong O $2p$--Ni $3d$ hybridization~\cite{zhang2026distinct}. For $\pi$ polarization, the pre-edge is shifted lower in energy by approximately 0.2~eV relative to $\sigma$ polarization. The Ni $L_3$-edge shows a corresponding energy shift, with two broad spectral components more clearly resolved for $\pi$ polarization and displaced together by approximately 0.2~eV to lower energy. The similar polarization-dependent shifts at both the Ni $L_3$ and O $K$ edges indicate a common anisotropy of the hybridized Ni--O states, while the slightly broader $\pi$-polarized profiles are consistent with interlayer-hybridized molecular orbitals involving inner-apical O $2p_z$ hopping~\cite{he2026layer,labollita2024electronic}. Extended XAS measurements for both Ni $L_{2,3}$-edges and a broader O $K$-edge energy range are reported in Supplemental Material Sec.~S3.

\begin{figure*}[t]
    \centering
    \includegraphics[width=\textwidth]{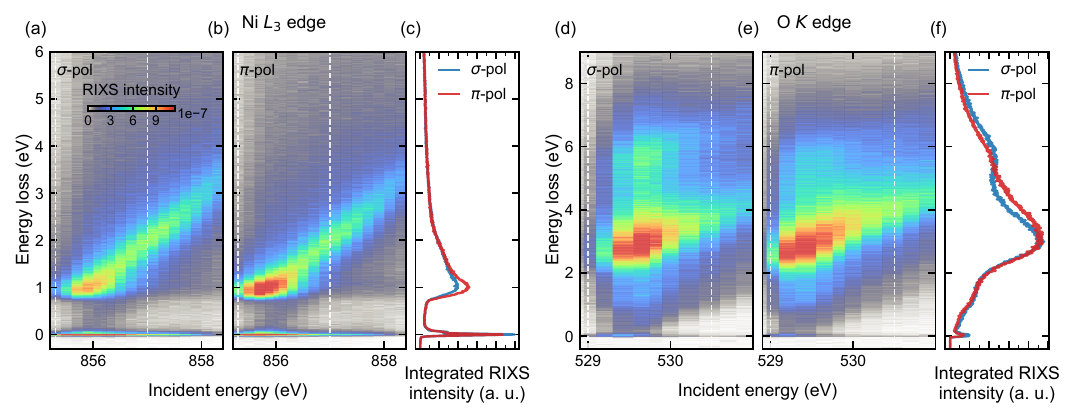}\vspace{-3mm}
    \caption{(Color online) Incident-energy and polarization dependence of the Ni $L_3$- and O $K$-edge RIXS response. (a),(b) Ni $L_3$-edge RIXS intensity as a function of incident energy and energy loss for $\sigma$- and $\pi$-polarized incident x-rays, respectively. (c) Corresponding energy-loss profiles integrated over the incident-energy ranges marked by the vertical dashed lines in (a) and (b). The broad $dd$ excitation manifold near 1~eV is enhanced for $\pi$ polarization, revealing a pronounced polarization dependence of the local Ni orbital excitations. (d),(e) O $K$-edge RIXS maps measured with $\sigma$- and $\pi$-polarized incident x-rays, respectively. (f) Corresponding incident-energy-integrated O $K$-edge RIXS profiles. Below approximately 3~eV, the response is similar for both polarizations, whereas the higher-energy charge-transfer sector shows a redistribution of spectral weight, indicating contributions from multiple oxygen-hybridization channels.}\vspace{-3mm}
    \label{fig:incident_energy_rixs}
\end{figure*}

We next examine the incident-energy dependence of the RIXS response across the Ni $L_3$ and O $K$ resonances in Fig.~\ref{fig:incident_energy_rixs}. Consistent with the XAS results, we find that the resonant incident energy shifts to lower energy for $\pi$ polarization. At both edges, the RIXS maps resolve characteristic electronic excitations of RP nickelates, including Raman-like $dd$ excitations centered near 1~eV, broader charge-transfer features above approximately 2~eV, and fluorescence-like spectral weight extending along the diagonal direction~\cite{chen2024electronic,tenhuisen2025magnetic,he2026layer}. The relative spectral weights differ between the two edges, as the Ni $L_3$ resonance [Figs.~\ref{fig:incident_energy_rixs}(a) and (b)] features local $dd$ excitations, whereas the O $K$ resonance [Figs.~\ref{fig:incident_energy_rixs}(d) and (e)] is more sensitive to the charge-transfer response~\cite{shen2022role,zhang2026distinct,he2026layer}. We observe little polarization dependence in the fluorescence-like continuum, consistent with its origin in more extended hybridized Ni--O states with comparable contributions from the $3d_{x^2-y^2}$ and $3d_{3z^2-r^2}$ orbitals~\cite{he2026layer}. By contrast, the incident-energy-integrated Ni $L_3$-edge spectra in Fig.~\ref{fig:incident_energy_rixs}(c) show a clear enhancement of the $dd$ manifold for the $\pi$ polarization, indicating a larger contribution from Ni $3d_{3z^2-r^2}$-derived states to the orbital-excitation response. We also find that the $dd$ excitation envelope in Pr$_4$Ni$_3$O$_{10}$ is relatively broad. This broadening may arise from a shorter lifetime of the $dd$ excitations, although contributions from the more distorted and inequivalent NiO$_6$ environments within the trilayer may also be present. Comparison with La$_4$Ni$_3$O$_{10}$ in Supplemental Material Sec.~S5 confirms the broader orbital-excitation response of the Pr compound and the absence of the broad 3--6~eV La-hybridized charge-transfer response.

At the O $K$ edge [Figs.~\ref{fig:incident_energy_rixs}(d) and (e)], we find the strongest polarization dependence in the charge-transfer sector. The integrated spectra in Fig.~\ref{fig:incident_energy_rixs}(f) remain similar below approximately 3~eV but diverge at higher energy loss. The $\sigma$-polarized response forms a sharper feature near 6~eV, whereas the $\pi$-polarized response extends over a broader energy range. These distinct line shapes indicate different distributions of charge-transfer final states for the predominantly planar and out-of-plane oxygen channels. We associate the broader $\pi$-polarized response with strong interlayer hybridization between inner-apical O $2p_z$ and Ni $3d_{3z^2-r^2}$ states, which splits the trilayer $3d_{3z^2-r^2}$-derived states into bonding, nonbonding, and antibonding molecular subbands and distributes the corresponding O $2p_z$ spectral weight over a wider energy range~\cite{labollita2024electronic,he2026layer}. The sharper $\sigma$-polarized feature instead reflects planar O $2p_{x,y}$--Ni $3d_{x^2-y^2}$ response distributed over a narrower energy range~\cite{he2026layer}. At much lower energy loss, not resolved on the scale of the maps shown above, we also observe a pronounced excitation near 50~meV, consistent with phonons previously detected by O $K$-edge RIXS in RP nickelates~\cite{he2026layer,zhang2026distinct}. This mode is resonantly enhanced near the O $K$ pre-edge and is most clearly visible for $\sigma$ polarization. Enlarged low-energy maps and further discussion are provided in Supplemental Material Sec.~S6.

Taken together, the polarization-dependent XAS and RIXS measurements reveal a clear anisotropy of the hybridized Ni--O states at both the Ni $L_3$ and O $K$ edges. The common polarization-dependent resonance shifts, enhanced $\pi$-polarized $dd$ response, and redistribution of the O $K$-edge charge-transfer spectral weight distinguish the predominantly planar Ni $3d_{x^2-y^2}$--O $2p_{x,y}$ and out-of-plane Ni $3d_{3z^2-r^2}$--O $2p_z$ channels. The absence of an overlapping La $M_4$ contribution in  Pr$_4$Ni$_3$O$_{10}$ allows these multiorbital features to be resolved directly.
Compared with La$_4$Ni$_3$O$_{10}$, the Pr compound exhibits a similar overall polarization-dependent Ni--O response, but with broader orbital excitations and without the La-hybridized 3--6~eV charge-transfer feature.
These results establish the electronic-structure context for the low-energy magnetic response discussed next.
\begin{figure*}[t]
    \centering
    \includegraphics[width=\textwidth]{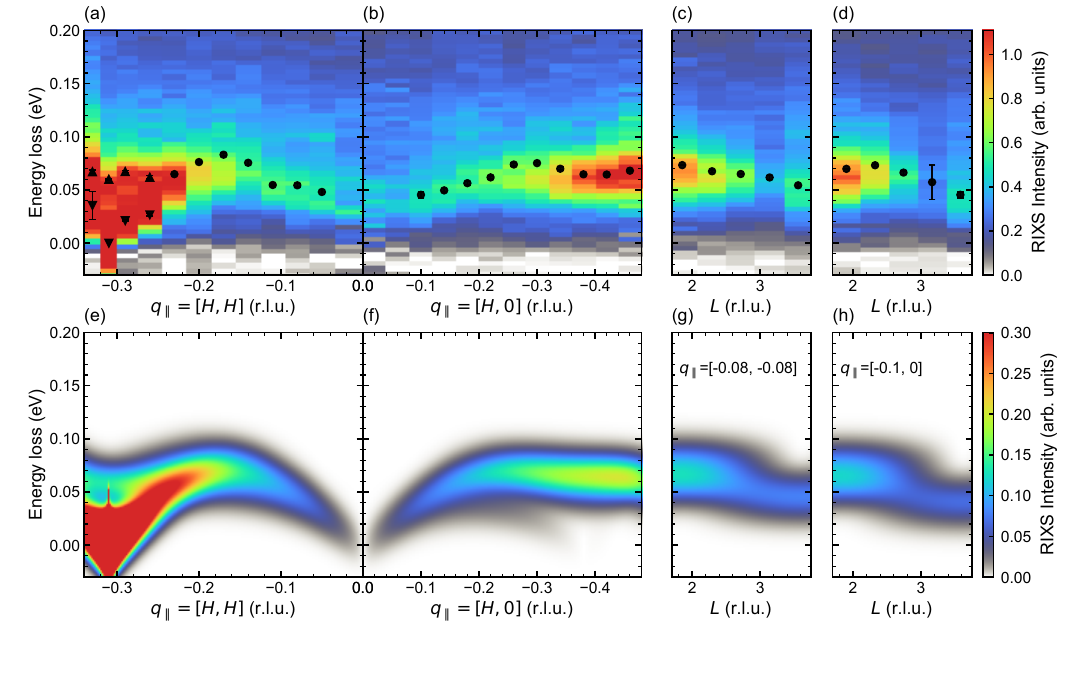}\vspace{-7mm}
    \caption{Magnetic-excitation dispersion and effective spin-wave model. (a)--(d) Elastic-line-subtracted Ni $L_3$-edge RIXS intensity along (a) $\mathbf{q}_{\parallel}=(H,H)$, (b) $\mathbf{q}_{\parallel}=(H,0)$, and the out-of-plane direction at fixed (c) $\mathbf{q}_{\parallel}=(-0.08,-0.08)$ and (d) $\mathbf{q}_{\parallel}=(-0.10,0)$. Symbols show excitation energies extracted from fits to the individual spectra. Different symbols denote the two low-energy branches where both are resolved. Along $(H,H)$, the excitations soften and spectral weight accumulates near the SDW wave vector $H\simeq-0.31$ r.l.u. (e)--(h) Corresponding dynamical magnetic response calculated using linear spin-wave theory for the effective outer-layer SDW model in Fig.~\ref{fig:structure_model}(b), evaluated along the experimental trajectories and convolved with a 30~meV FWHM Gaussian. Experimental spectra were measured with $\pi$-polarized incident x-rays tuned to the magnetic resonance near 855.8~eV.}\vspace{-4mm}
    \label{fig:qmap_lswt}
\end{figure*}

\vspace{-3mm}
\section{Magnetic excitations and interlayer coupling}\vspace{-3mm}

We now examine the momentum dependence of the low-energy Ni $L_3$-edge RIXS response [Figs.~\ref{fig:qmap_lswt}(a)--(d)]. Using $\pi$-polarized incident x-rays tuned to the first Ni $L_3$ resonance near 855.8~eV, we resolve dispersive low-energy excitations along both nominal in-plane directions $(H,H)$ and $(H,0)$, with fitted energies spanning approximately 20--85~meV. Their magnetic assignment is supported by polarimetric RIXS measurements on related RP nickelates~\cite{chen2024electronic,chen2026dissecting,chan2026collective}. The most pronounced evolution occurs along $(H,H)$, where the excitation softens as $H$ approaches $-0.31$ r.l.u. and spectral weight accumulates near zero energy loss. This momentum coincides with the incommensurate SDW wave vector established by diffraction, directly connecting the dispersive excitation to the density-wave magnetic state~\cite{zhang2020intertwined,samarakoon2023bootstrapped}. The spectra also contain a broad mid-infrared (MIR) continuum extending to several hundred meV before merging into the $dd$ excitations. The MIR response resonates at the Ni $L_3$ edge and varies only weakly with momentum, consistent with a resonantly enhanced metallic Ni--O background. Its weaker intensity in Pr$_4$Ni$_3$O$_{10}$ relative to La$_4$Ni$_3$O$_{10}$, consistent with the lower metallic conductivity of the Pr compound, makes the low-energy magnetic excitations more clearly resolved~\cite{chen2024electronic,chen2026dissecting,chan2026collective}. A direct comparison is provided in Supplemental Material Sec.~S5.

At most momenta, we resolve one or two magnetic excitations. Where the response is weaker or approaches the elastic line, we detect a broad low-energy feature and cannot separate individual modes. We extract the excitation energies by fitting the spectra with an elastic line, a smooth MIR contribution, and damped magnetic-excitation components. For clarity, Fig.~\ref{fig:qmap_lswt} shows the intensity maps after subtracting the fitted elastic contribution; complete spectral decompositions and fitting details are provided in Supplemental Material Sec.~S9. The $L$ scans reveal a pronounced modulation of the magnetic spectral weight, while the fitted excitation energies vary more modestly. Models based on magnetic diffraction suggest a configuration with antiparallel outer-layer moments and the inner layer near a node of the SDW modulation~\cite{zhang2020intertwined, samarakoon2023bootstrapped}. The observed periodicity of the magnetic excitations here is consistent with this ground state spin configuration. By contrast, introducing a sizable inner-layer moment would instead produce an up--down--up configuration and double the spectral-weight modulation periodicity along $L$, inconsistent with the data~\cite{he2026layer}.

To quantify the interactions underlying the measured dispersion, we model the response using linear spin-wave theory for an effective outer-layer SDW model~\cite{chen2024electronic,chen2026dissecting,chan2026collective,he2026layer}, implemented using the \textsc{Sunny} package~\cite{dahlbom2025sunny}. The in-plane interactions $J_1$, $J_2$, and $J_3$ couple first-, second-, and third-nearest-neighbor sites within the two outer NiO$_2$ layers, whose ordered moments are opposite in phase. We therefore use $J_z$ to parameterize the effective coupling between the two outer-layer SDW components across the trilayer. Pr$_4$Ni$_3$O$_{10}$ also contains two orthogonal structural-twin domains that contribute to spatially averaged x-ray measurements~\cite{jiang2026direct}. Following Ref.~\cite{he2026layer}, we therefore calculate the nominal $(H,H)$ response as an equal-weight incoherent average of the two twins, with SDW propagation vectors along $(H,H)$ and $(H,-H)$.

A simultaneous fit to the two nominal in-plane scans and two fixed-$q_{\parallel}$ $L$ scans gives $J_1=15.7(3)$~meV, $J_2=1.7(3)$~meV, $J_3=9.8(1)$~meV, and $J_z=15.8(6)$~meV. The in-plane exchange couplings are constrained in the fitting to correctly capture the incommensurate SDW wavevector (Supplemental Material Sec.~S10 \cite{supp}). The calculated response [Figs.~\ref{fig:qmap_lswt}(e)--(h)] reproduces both the magnetic-excitation dispersion and the relative intensity distribution along both in-plane and out-of-plane trajectories. Most notably, the effective coupling $J_z$ is comparable to, and slightly larger than, the nearest-neighbor in-plane coupling $J_1$, despite spanning the full trilayer across the middle NiO$_2$ layer. This unusually large out-of-plane interaction is consistent with the strong Ni $3d_{3z^2-r^2}$--O $2p_z$ hybridization resolved via XAS, suggesting that the substantial participation of this out-of-plane orbital channel supports magnetic coupling across the trilayer. The importance of this pathway is reinforced by comparison with square-planar R$_4$Ni$_3$O$_8$, where the absence of apical oxygen suppresses out-of-plane hybridization and the dominant magnetic exchange remains within the NiO$_2$ planes, supporting an important role for apical-oxygen-mediated hybridization in the sizable out-of-plane magnetic interaction of Pr$_4$Ni$_3$O$_{10}$~\cite{lin2021strong}.

The comparable in-plane and out-of-plane interaction scales place the trilayer near an intermediate-coupling regime where conventional spin-wave theory is not fully controlled~\cite{ganesh2011neel,lohofer2015dynamical}. We therefore further test the robustness of the extracted interlayer coupling using complementary entangled-dimer calculations, together with a stripe-dimer calculation described in Supplemental Material Sec.~S11.
These alternative descriptions yield $J_z\simeq15$--21~meV, comparable to the leading in-plane interactions and consistent with the linear-spin-wave result. 
In this intermediate-coupling regime, neither an expansion about ordered moments, as in conventional spin-wave theory, nor a strong-dimer expansion is fully controlled; nevertheless, both consistently require a substantial effective interlayer coupling.
Recent RIXS studies of La$_4$Ni$_3$O$_{10}$ find similar interaction scales, with leading in-plane interactions of $\sim10$--16~meV and effective interlayer couplings of $\sim20$--22~meV~\cite{chen2026dissecting,chan2026collective,he2026layer}. The interlayer interaction is therefore similar between Pr- and La-based trilayer compounds, in sharp contrast to bilayer La$_3$Ni$_2$O$_7$, where it reaches approximately 50--70~meV and substantially exceeds the in-plane interactions~\cite{xie2024strong,chen2026nature,chen2024electronic,he2026layer}.

More broadly, although the local-moment descriptions above provide good descriptions of the observed dispersions, the metallic nature of these materials suggests that  charge degrees of freedom may also play a role in the magnetism in trilayer nickelates. The fitted exchange parameters therefore represent effective interaction scales incorporating the combined effect of local moments together with electronic correlations and itinerant band effects. 
Recent multiband Hartree-Fock and random-phase approximation (RPA) calculations reproduce the outer-layer-dominated SDW structure and spin-wave-like collective excitations within an itinerant framework, without assuming preexisting localized moments~\cite{mei2026itinerant,zhan2026magnetic,zhang2025sdw}. However, extending such self-consistent calculations to the experimentally observed incommensurate state of trilayer nickelates such as Pr$_4$Ni$_3$O$_{10}$, including its charge modulation and full trilayer orbital structure, would require a substantially larger magnetic unit cell than the commensurate approximations used so far.

\section{Conclusions}\vspace{-3mm}

In summary, we probed the multiorbital electronic structure and magnetic excitation spectrum of Pr$_4$Ni$_3$O$_{10}$ using XAS and RIXS. Polarization-dependent measurements distinguish planar and out-of-plane Ni--O hybridization channels, while momentum-resolved RIXS constrains the magnetic interactions within the trilayer and reveals the influence of the incommensurate SDW state on the collective magnetic response. A comparison with La$_4$Ni$_3$O$_{10}$ shows that the dominant magnetic interaction scales of trilayer RP nickelates remain robust against rare-earth substitution despite substantial changes in structural distortion and metallic response. 
The contrast between this weak La--Pr dependence and the pronounced variation of the interaction scales across different layer numbers identifies the multilayer architecture as a primary control parameter of the magnetic-interaction hierarchy in RP nickelates. The interaction scales established here provide an ambient-pressure reference for determining how this hierarchy evolves as pressure suppresses the density-wave state and drives Pr$_4$Ni$_3$O$_{10}$ toward superconductivity~\cite{huang2024signature,zhang2025bulk,chen2025low,pei2026unveiling}, and for testing proposals linking interlayer coupling to superconducting pairing~\cite{lu2024interlayer,huang2024interlayer}.


\section{Data Availability}\vspace{-3mm}

The supporting data for this article are openly available from the Zenodo database \cite{repo}. The raw data collected at ESRF are archived under DOI \href{https://doi.esrf.fr/10.15151/ESRF-ES-2227853331}{10.15151/ESRF-ES-2227853331} and will become openly accessible following the three-year embargo period.

\begin{acknowledgments}
We thank Yifeng Zhao for insightful discussions. We acknowledge the European Synchrotron Radiation Facility (ESRF) for provision of synchrotron radiation facilities under proposal HC6368 on beamline ID32; we gratefully thank Céline Allais for her technical assistance and rapid support for the vacuum intervention during the beamtime.
Work by Z.G., S.F.R.T., X.G., W.H., M.P.M.D., and M.M. was supported by the U.S. Department of Energy (DOE), Division of Materials Science, under Contract No. DE-SC0012704.  S.F.R.T. acknowledges additional support from the DOE, Office of Science, Office of Workforce Development for Teachers and Scientists, Office of Science Graduate Student Research (SCGSR) program. The SCGSR program is administered by the Oak Ridge Institute for Science and Education for the DOE under Contract No. DE-SC0014664. W.H. acknowledges additional support by the U.S. Department of Energy, Office of Science, Basic Energy Sciences, Materials Sciences and Engineering Division, under contract DE-AC02-76SF00515. Work by X.C., H.Z., and J.F.M. in the Materials Science Division at Argonne National Laboratory (crystal growth) was supported by the U.S. Department of Energy Office of Science, Basic Energy Sciences, Materials Science and Engineering Division. 
\end{acknowledgments}

\nocite{bisogni2016ground, dallera2001charge, fabbris2023resonant}

\bibliography{ref} 
\end{document}


\title{Supplemental Material for: Multiorbital Anisotropy and Magnetic Excitations in Trilayer Nickelate Pr$_4$Ni$_3$O$_{10}$}

\author{Ziqiang Guan}
\thanks{these authors contributed equally}
\affiliation{Department of Physics, Harvard University, Cambridge, Massachusetts 02138, USA}

\author{Sophia F. R. TenHuisen}
\thanks{these authors contributed equally}
\affiliation{Department of Physics, Harvard University, Cambridge, Massachusetts 02138, USA}
\affiliation{John A. Paulson School of Engineering and Applied Sciences, Harvard University, Cambridge, Massachusetts 02138, USA}

\author{Xuefei Guo}
\affiliation{Condensed Matter Physics and Materials Science Department, Brookhaven National Laboratory, Upton, New York 11973, USA}

\author{Wei He}
\affiliation{Condensed Matter Physics and Materials Science Department, Brookhaven National Laboratory, Upton, New York 11973, USA}
\affiliation{Stanford Institute for Materials and Energy Sciences,
SLAC National Accelerator Laboratory, Menlo Park, CA 94025, USA}

\author{Aulden K. Jones}
\affiliation{Department of Physics, Harvard University, Cambridge, Massachusetts 02138, USA}

\author{Marli dos Reis Cantarino}
\affiliation{European Synchrotron Radiation Facility, 71 Avenue des Martyrs, CS 40220, 38043 Grenoble Cedex 9, France}

\author{Xinglong Chen}
\author{Hong Zheng}
\author{J. F. Mitchell}
\affiliation{Materials Science Division, Argonne National Laboratory, Lemont, Illinois 60439, USA}

\author{Nicholas B. Brookes}
\affiliation{European Synchrotron Radiation Facility, 71 Avenue des Martyrs, CS 40220, 38043 Grenoble Cedex 9, France}

\author{Mark P. M. Dean}
\affiliation{Condensed Matter Physics and Materials Science Department, Brookhaven National Laboratory, Upton, New York 11973, USA}

\author{Matteo Mitrano}
\email{mmitrano@g.harvard.edu}
\affiliation{Department of Physics, Harvard University, Cambridge, Massachusetts 02138, USA}


\date{\today}
\maketitle
\vspace{-7mm}
\tableofcontents

\setstretch{1.3}

\section{Sample Preparation} \vspace{-3mm}

\begin{figure}[h]
    \centering
    \includegraphics[width=0.6\textwidth]{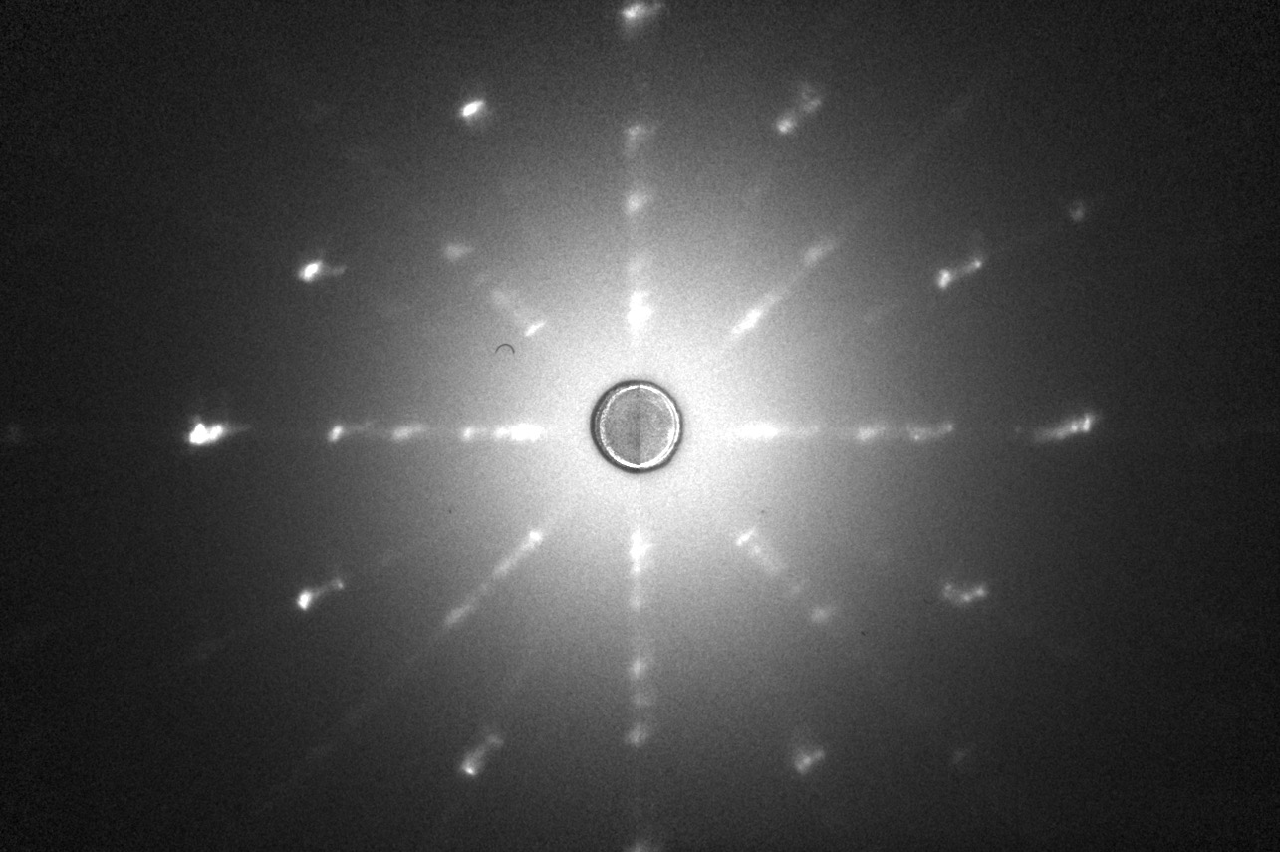}
    \caption{X-ray Laue pattern taken on the (001) surface of the Pr$_4$Ni$_3$O$_{10}$ single crystal used in this study.}\vspace{-3mm}
    \label{fig:S1_Pr4310_Laue}
\end{figure}

High-quality single crystals of Pr$_4$Ni$_3$O$_{10}$ were grown in a vertical optical-image floating-zone furnace under high O$_2$ pressure, following the procedure described in Ref.~\cite{zhang2020high}. At ambient pressure, Pr$_4$Ni$_3$O$_{10}$ crystallizes in the monoclinic $P2_1/a$ space group. Throughout this work, however, we adopt a pseudo-tetragonal unit-cell convention and express momentum transfer in reciprocal lattice units as $\mathbf{q}=(H,K,L)$, with lattice parameters $a=b=3.829$~\AA\ and $c=27.5$~\AA. Prior to the XAS and RIXS measurements, we verified crystal quality and orientation by x-ray Laue diffraction. The Laue pattern obtained from the $(001)$ surface is shown in Fig.~\ref{fig:S1_Pr4310_Laue}.

\vspace{-3mm}

\section{XAS and RIXS measurements}\vspace{-3mm}
\label{sec:method}

X-ray absorption spectroscopy (XAS) and high-energy-resolution resonant inelastic x-ray scattering (RIXS) measurements were performed at the ID32 beamline of the European Synchrotron Radiation Facility (ESRF)~\cite{brookes2018beamline}. After mounting the sample on the four-circle goniometer, its crystallographic orientation was refined \textit{in situ} using the $(002)$ and $(013)$ Bragg reflections, measured with the fundamental soft-x-ray beam and its second harmonic, respectively. All XAS and RIXS measurements were performed at 30~K, well below the SDW transition temperature, but above the $\sim26$~K transition coupling Pr and Ni magnetic moments~\cite{samarakoon2023bootstrapped}.

XAS spectra at the Ni $L_{2,3}$ and O $K$ edges were collected simultaneously in total-electron-yield (TEY) and total-fluorescence-yield (TFY) modes. The TFY signal was measured with a Si photodiode positioned at a scattering angle of $2\Theta=160^\circ$. Polarization-dependent XAS measurements were performed with both $\sigma$- and $\pi$-polarized incident x-rays ($\theta=15^\circ$), providing sensitivity predominantly to in- and out-of-plane unoccupied electronic states, respectively.

We measured XAS spectra by following two incident-energy acquisition protocols. High-resolution spectra over selected energy windows were acquired using point-by-point scans, with the monochromator stabilized at each prescribed incident energy before recording the detector signals. These scans provided the more reliable relative incident-energy calibration and were used to resolve the Ni $L_3$ and $L_2$ absorption edges and the O $K$-edge pre-edge region. Complementary continuous scans were acquired while sweeping the monochromator through the selected energy range without stopping at individual energy points. Their substantially shorter acquisition time enabled measurements over broader energy windows, encompassing the full Ni $L_{2,3}$-edge region for comparison of the relative spectral weights of the two edges and an extended range across the O $K$ edge.

For the RIXS measurements, the spectrometer arm was fixed at a scattering angle of $2\Theta=149.5^\circ$, except for the out-of-plane momentum-dependent scans. Incident-energy-dependent RIXS measurements across the Ni $L_3$ edge ($855.3$--$858.3$~eV) and O $K$ edge ($528.85$--$530.85$~eV) were performed with both $\sigma$- and $\pi$-polarized incident x-rays in the $(H,H,L)$ scattering plane at an incidence angle of $\theta=15^\circ$. In-plane and out-of-plane momentum-dependent spectra were collected with $\pi$-polarized incident photons at the magnon resonance energy of approximately $855.8$~eV.

The overall RIXS energy resolution was determined from the full width at half maximum (FWHM) of the elastic-scattering profile measured from amorphous carbon. Momentum-dependent Ni $L_3$-edge measurements were performed with a narrower monochromator exit slit, yielding an overall energy resolution of 24.5~meV. For the incident-energy-dependent measurements, a wider exit slit was used to increase the incident photon flux and reduce acquisition times. The resulting broader incident-photon bandwidth yielded overall energy resolutions of 26~meV at the Ni $L_3$ edge and 34~meV at the O $K$ edge. All RIXS spectra were normalized to the incident photon flux. For selected fits, we also normalized the spectra to the integrated $dd$-excitation intensity to consistently compare the low-energy response.

\vspace{-3mm}
\section{Polarization-Dependent XAS of \texorpdfstring{$\mathrm{Pr}_4\mathrm{Ni}_3\mathrm{O}_{10}$}{Pr4Ni3O10}}\vspace{-3mm}
\label{sec:XAS_full}

\begin{figure}
    \centering
    \includegraphics[width=0.7\textwidth]{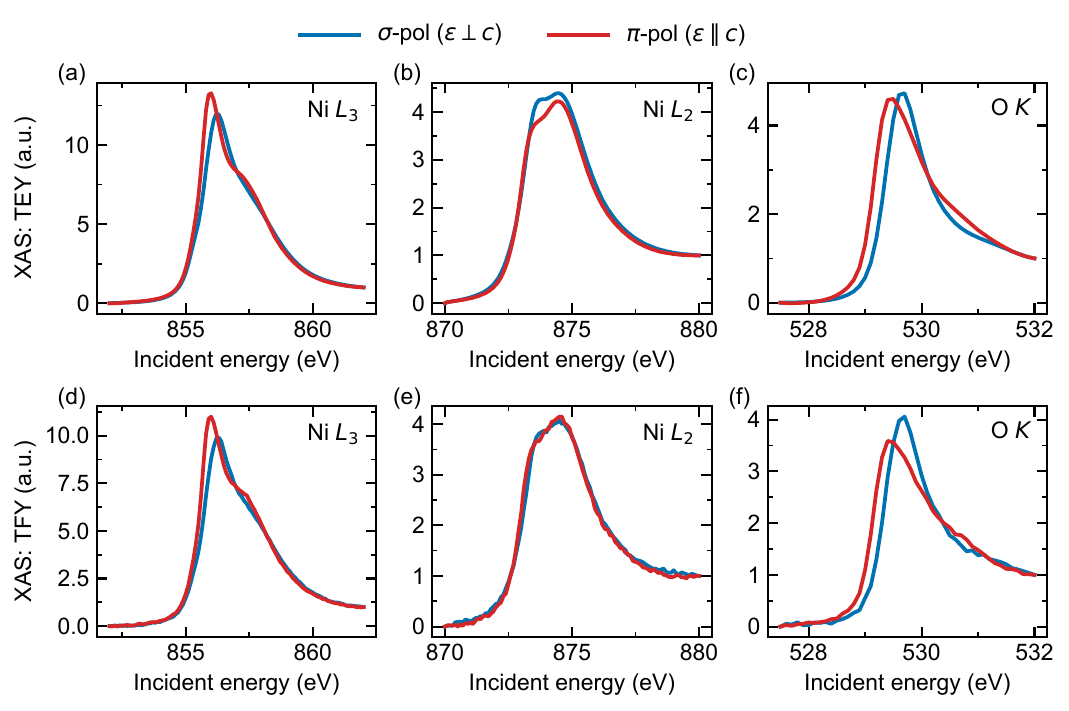}
    \caption{Polarization-dependent XAS spectra of Pr$_4$Ni$_3$O$_{10}$ near the Ni $L_3$-, Ni $L_2$-, and O $K$-edges. (a)--(c) TEY spectra and (d)--(f) TFY spectra measured with $\sigma$ polarization ($\boldsymbol{\varepsilon}\perp c$) and $\pi$ polarization ($\boldsymbol{\varepsilon}$ predominantly parallel to $c$ in the grazing-incidence geometry). The spectra were acquired using point-by-point incident-energy scans, normalized to the incident photon flux, and subsequently rescaled by setting the intensities at the low- and high-energy endpoints of each measured range to zero and unity, respectively.}
    \label{fig:S2_XAS_ascans}
\end{figure}

\begin{figure}
    \centering
    \includegraphics[width=0.7\textwidth]{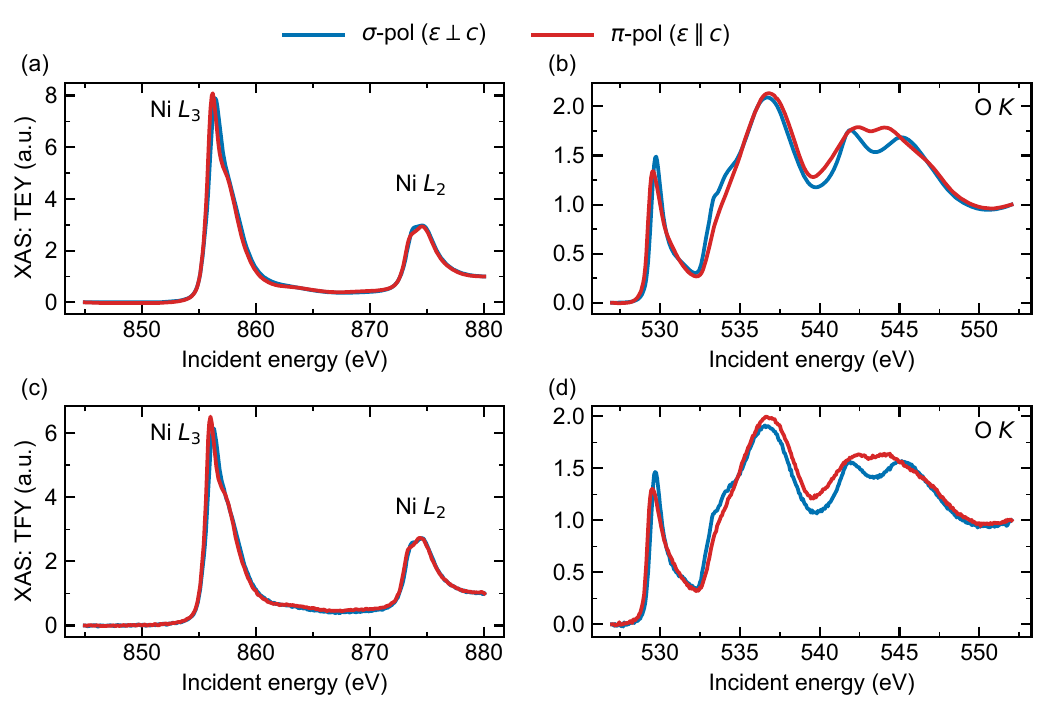}
    \caption{Extended-range polarization-dependent XAS spectra of Pr$_4$Ni$_3$O$_{10}$. (a),(b) TEY spectra and (c),(d) TFY spectra measured across the Ni $L_{2,3}$- and O $K$-edges, respectively. Blue and red curves show measurements with $\sigma$ polarization ($\boldsymbol{\varepsilon}\perp c$) and $\pi$ polarization ($\boldsymbol{\varepsilon}$ with a dominant $c$-axis component), respectively. We acquired the spectra using continuous incident-energy scans and normalized each spectrum by setting the intensities at the low- and high-energy endpoints of the measured range to zero and unity, respectively. The extended scans capture both the Ni $L_3$ and $L_2$ edges within a common energy range and reveal higher-energy O $K$-edge structures beyond the pre-edge features shown in Fig.~\ref{fig:S2_XAS_ascans}.}
    \label{fig:S3_XAS_escans}
\end{figure}

We measured polarization-dependent XAS at the Ni $L_{2,3}$ and O $K$ edges using both TEY and TFY detection. In TEY, we recorded the sample drain current generated by emitted electrons, making the signal predominantly sensitive to the near-surface region because of their short escape depth. In TFY, we detected the emitted x-ray fluorescence and therefore probed substantially deeper into the bulk, although saturation and self-absorption can modify relative peak intensities. Because TEY is more surface sensitive, it is also more susceptible to contamination, degradation, and reconstruction. As shown in Figs.~\ref{fig:S2_XAS_ascans} and \ref{fig:S3_XAS_escans}, we observe consistent absorption features and polarization dependence in both detection channels, indicating that the near-surface electronic structure is broadly representative of the bulk and showing no evidence of pronounced surface degradation or of a distinct surface phase.

We measured both $\sigma$ and $\pi$ incident polarizations under grazing-incidence conditions. The electric field of the $\sigma$-polarized beam lies within the crystallographic $ab$ plane, whereas the $\pi$-polarized beam has a dominant $c$-axis component, providing sensitivity to in- and out-of-plane unoccupied electronic states, respectively. To compare the two polarizations, we normalized each spectrum using the intensities at the low- and high-energy endpoints of the measured range, setting the former to zero and the latter to unity: $I_{\mathrm{norm}}(E)=\frac{I(E)-I(E_{\mathrm{left}})}{I(E_{\mathrm{right}})-I(E_{\mathrm{left}})}$. This normalization assumes negligible resonant absorption sufficiently below the edge and a polarization-independent absorption level sufficiently above it. It removes differences in overall offset and scale while preserving the polarization-dependent line shape within each energy window.

Figure~\ref{fig:S2_XAS_ascans} shows point-by-point incident-energy scans near the Ni $L_3$-, Ni $L_2$-, and O $K$-edges, with TEY and TFY spectra in the upper and lower rows, respectively. We observe a clear polarization dependence at the Ni $L_3$-edge and in the O $K$- pre-edge region, whereas the Ni $L_2$ response is less anisotropic. The Ni $L_3$ spectra contain a lower-energy main resonance and a broader higher-energy feature with stronger ligand-hole and Ni--O-hybridized character. Both features are shifted to lower incident energy for $\pi$ polarization by approximately 0.2~eV. The higher-energy component is also broader for $\sigma$ polarization, indicating a polarization-dependent distribution of the hybridized unoccupied states. We observe a comparable shift of approximately 0.2~eV at the O $K$ pre-edge. Together, the Ni $L_3$- and O $K$-edge spectra reveal a clear energetic and line-shape anisotropy between the predominantly in-plane and out-of-plane components of the hybridized Ni--O electronic structure. We complemented the point-by-point measurements with continuous incident-energy scans over broader energy ranges, shown in Fig.~\ref{fig:S3_XAS_escans}. At the Ni $L$ edge, these scans cover the full $L_{2,3}$ region and allow direct comparison of the relative edge intensities and their polarization dependence. At the O $K$ edge, they reveal higher-energy absorption features beyond the pre-edge region resolved in the point-by-point scans. The continuous scans thus provide an overview of the full Ni $L_{2,3}$- and O $K$-edge responses, while the point-by-point measurements provide more accurately calibrated spectra near the absorption edges.

\section{\texorpdfstring{$\mathrm{Ni}$}{Ni} $L_3$-Edge XAS Comparison between
\texorpdfstring{$\mathrm{Pr}_4\mathrm{Ni}_3\mathrm{O}_{10}$}{Pr4Ni3O10}
and
\texorpdfstring{$\mathrm{La}_4\mathrm{Ni}_3\mathrm{O}_{10}$}{La4Ni3O10}}
\label{sec:XAS_PrvsLa}\vspace{-3mm}

\begin{figure}
    \centering
    \includegraphics[width=\textwidth]{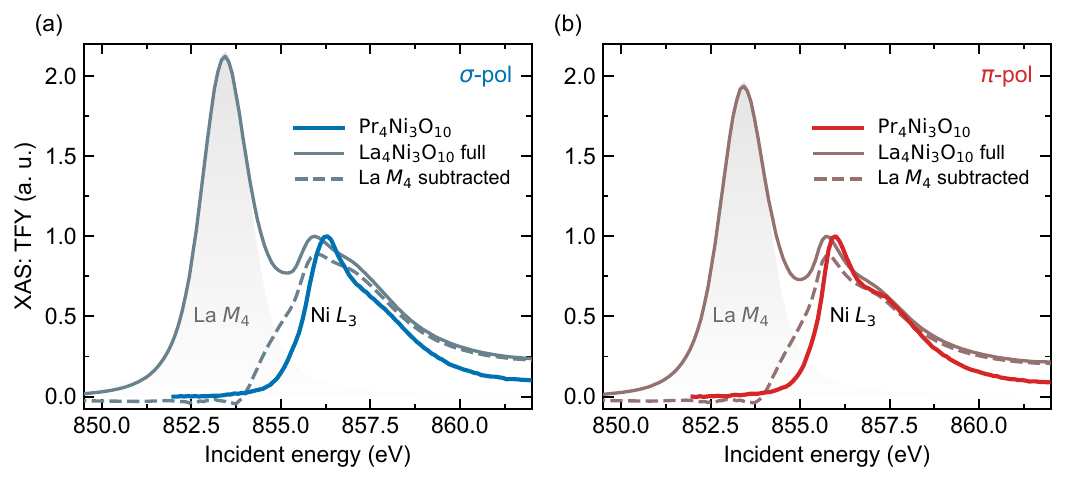}
    \caption{Comparison of the Ni $L_3$-edge TFY-XAS spectra of Pr$_4$Ni$_3$O$_{10}$ and La$_4$Ni$_3$O$_{10}$ for (a) $\sigma$ and (b) $\pi$ incident polarizations. Blue and red curves show the Pr$_4$Ni$_3$O$_{10}$ spectra for $\sigma$ and $\pi$ polarization, respectively. For La$_4$Ni$_3$O$_{10}$, solid gray curves show the measured spectra, gray shaded regions the fitted La $M_4$ contribution, and dashed gray curves the spectra after its subtraction. We shifted the La$_4$Ni$_3$O$_{10}$ spectra by a constant energy offset to align the Ni $L_3$ resonances and rescaled them so that the Ni $L_3$ peak height approximately matched that of Pr$_4$Ni$_3$O$_{10}$. Reference XAS spectra of La$_4$Ni$_3$O$_{10}$ are reproduced from Ref.~\cite{he2026layer}.}
    \label{fig:S4_Pr4310_La4310}
\end{figure}

Figure~\ref{fig:S4_Pr4310_La4310} compares the Ni $L_3$-edge XAS of Pr$_4$Ni$_3$O$_{10}$ and La$_4$Ni$_3$O$_{10}$. All spectra were acquired in TFY mode using the point-by-point incident-energy protocol described in Secs.~\ref{sec:method} and \ref{sec:XAS_full}. We measured Pr$_4$Ni$_3$O$_{10}$ at the ID32 beamline of ESRF, while the La$_4$Ni$_3$O$_{10}$ reference spectra were acquired at the SIX beamline of the National Synchrotron Light Source II. Because the datasets were collected at different beamlines and under different experimental geometries, we use this comparison primarily to assess the relative Ni $L_3$ line shapes rather than absolute intensities or polarization-resolved spectral weights. In La$_4$Ni$_3$O$_{10}$, the low-energy side of the Ni $L_3$-edge strongly overlaps with the La $M_4$-edge. To estimate this contribution, we fitted the spectrum below the onset of the Ni $L_3$ resonance, where La $M_4$ absorption dominates, using a Voigt profile and a constant offset
\begin{equation}
I_{\mathrm{La}\,M_4}(E)
=
A\,
\frac{V(E-E_0;\sigma,\gamma)}
     {V(0;\sigma,\gamma)}
+C,
\end{equation}
where $A$ is the peak amplitude, $E_0$ the peak center, $\sigma$ and $\gamma$ the Gaussian and Lorentzian widths, respectively, and $C$ a constant offset. The normalization by $V(0;\sigma,\gamma)$ makes $A$ equal to the fit peak height above the offset. We then extrapolated the fitted La $M_4$ contribution through the overlapping region and subtracted it from the La$_4$Ni$_3$O$_{10}$ spectrum. In Fig.~\ref{fig:S4_Pr4310_La4310}, solid gray curves show the full La$_4$Ni$_3$O$_{10}$ spectra, gray shading the fitted La $M_4$ contribution, and dashed gray curves the spectra after subtraction. Because the La $M_4$ line shape strongly overlaps the Ni $L_3$ resonance and is not fully captured by the single Voigt profile used here, its contribution cannot be modeled and isolated perfectly. Therefore, the weak shoulders remaining on the low-energy side of the subtracted spectra are most likely residuals from the La $M_4$ subtraction, not intrinsic Ni $L_3$ response. This difficulty further highlights the advantage of Pr$_4$Ni$_3$O$_{10}$, whose Ni $L_3$ resonance is not obscured by an overlapping rare-earth absorption edge. For comparison with Pr$_4$Ni$_3$O$_{10}$, we shifted the La$_4$Ni$_3$O$_{10}$ spectra by a constant energy offset to align the Ni $L_3$ resonance and rescaled them to approximately match the Ni $L_3$ peak height. These adjustments account for differences in absolute energy calibration and overall signal scale between the two beamlines and carry no physical significance. The two compounds exhibit broadly similar Ni $L_3$ line shapes, although the Pr$_4$Ni$_3$O$_{10}$ resonance appears somewhat narrower. 

\section{Overview and Comparison of the \texorpdfstring{$\mathrm{Ni}$}{Ni} $L_3$- and O $K$-Edge RIXS Spectra}
\label{sec:RIXS_comparison}

\begin{figure}
    \centering
    \includegraphics[width=0.8\textwidth]{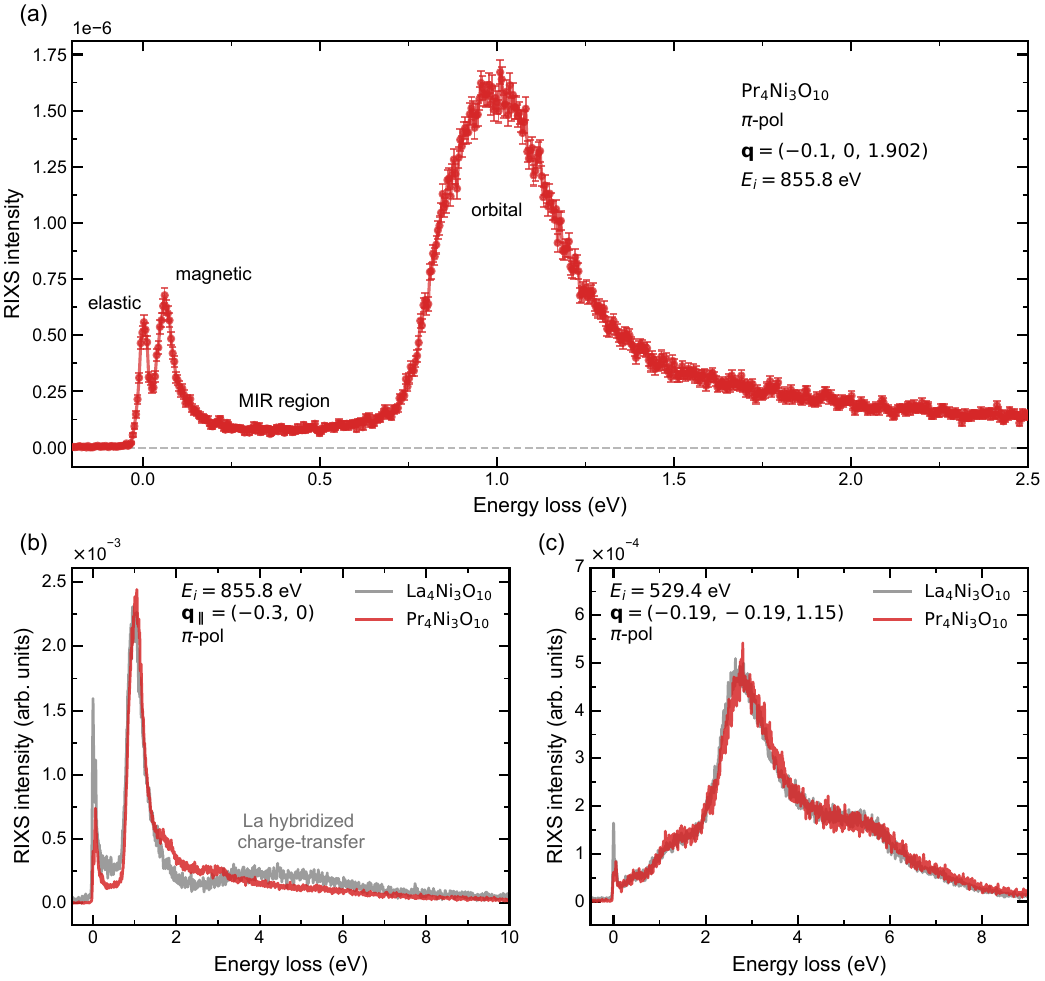}
    \caption{Overview and rare-earth dependence of the RIXS response. (a) Representative Ni $L_3$-edge RIXS spectrum of Pr$_4$Ni$_3$O$_{10}$ measured with $\pi$-polarized incident x-rays at $\mathbf{q}=(-0.10,0,1.902)$ r.l.u. and $E_i=855.8$~eV. The labeled contributions include quasi-elastic scattering near zero energy loss, the low-energy magnetic excitation at 40--70~meV, the broad mid-infrared (MIR) continuum, and orbital excitations near and above 1~eV. Error bars represent statistical uncertainties. (b) Ni $L_3$-edge RIXS spectra of La$_4$Ni$_3$O$_{10}$ and Pr$_4$Ni$_3$O$_{10}$ measured with the same incident polarization and in-plane momentum transfer, $\mathbf{q}_{\parallel}=(-0.3,0)$ r.l.u. Each spectrum is normalized to the integrated $dd$-excitation intensity between 0.7 and 1.4~eV. The broad feature between approximately 3 and 6~eV in La$_4$Ni$_3$O$_{10}$ arises from a La-hybridized charge-transfer response overlapping the Ni $L_3$-edge response. (c) O $K$-edge RIXS spectra measured at the same incident angle, aligned resonant incident energy, and $\pi$ polarization. Each spectrum is normalized to the integrated charge-transfer intensity between 1 and 5~eV. Reference spectra for La$_4$Ni$_3$O$_{10}$ are reproduced from Ref.~\cite{he2026layer}.}
    \label{fig:S5_RIXS_example}
\end{figure}

A representative Ni $L_3$-edge RIXS spectrum of Pr$_4$Ni$_3$O$_{10}$ is shown in Fig.~\ref{fig:S5_RIXS_example}(a), measured with $\pi$-polarized incident x-rays at $\mathbf{q}=(-0.10,0,1.902)$ r.l.u. and $E_i=855.8$~eV. Quasi-elastic scattering appears near zero energy loss, followed by a magnetic excitation at 40--70~meV, a broad mid-infrared (MIR) continuum, and orbital excitations near and above 1~eV. The broad orbital-excitation envelope is consistent with a high-spin $3d^8\underline{L}$ configuration, where $\underline{L}$ denotes a hole in predominantly ligand-centered states~\cite{bisogni2016ground,tenhuisen2025magnetic}. As shown by the incident-energy-dependent RIXS maps in the main text, the MIR spectral weight becomes appreciable only near the Ni $L_3$ resonance, linking it to the intrinsic Ni--O electronic response rather than an incident-energy-independent background. Its broad line shape is consistent with a resonantly enhanced metallic background similar to that reported in other metallic Ruddlesden--Popper nickelates~\cite{tenhuisen2025magnetic}.

Figure~\ref{fig:S5_RIXS_example}(b) compares the Ni $L_3$-edge spectra of La$_4$Ni$_3$O$_{10}$ and Pr$_4$Ni$_3$O$_{10}$ at the same in-plane momentum transfer and incident polarization. To compare the two datasets, we normalize each spectrum to the integrated $dd$-excitation intensity between 0.7 and 1.4~eV. After normalization, La$_4$Ni$_3$O$_{10}$ exhibits stronger relative MIR spectral weight, qualitatively consistent with its stronger metallic response and the lower conductivity of Pr$_4$Ni$_3$O$_{10}$ reported in transport measurements~\cite{rout2020structural,huangfu2020anisotropic}. The most prominent high-energy difference is the broad enhancement between approximately 3 and 6~eV in La$_4$Ni$_3$O$_{10}$. 
Although charge-transfer excitations are known to occur in La compounds under direct La $M$-edge excitation~\cite{dallera2001charge}, incident-energy-dependent RIXS measurements of La$_4$Ni$_3$O$_{10}$ show that this feature is enhanced near the Ni $L_3$ resonance but does not peak at the La $M_4$ resonance itself~\cite{fabbris2023resonant}. This incident-energy dependence argues against a simple assignment to fluorescence or charge-transfer excitations originating directly from the La $M_4$ edge. The presence of the feature in La$_4$Ni$_3$O$_{10}$ but not in Pr$_4$Ni$_3$O$_{10}$ instead suggests a charge-transfer excitation involving La-hybridized states, potentially with mixed La, O, and Ni orbital character. We therefore refer to this feature as a La-hybridized charge-transfer response.

The $dd$-excitation feature near 1--2~eV is slightly broader in Pr$_4$Ni$_3$O$_{10}$ than in La$_4$Ni$_3$O$_{10}$. This difference may reflect a shorter lifetime of the $dd$ excitations in the Pr compound. Structural effects may also contribute, since the smaller Pr ion induces stronger distortions of the NiO$_6$ octahedra, while the inner- and outer-layer Ni sites experience distinct local coordination environments~\cite{rout2020structural,huangfu2020anisotropic}. These differences could modify the energies and linewidths of the orbital excitations and contribute to the broader unresolved $dd$ envelope observed in Pr$_4$Ni$_3$O$_{10}$. The quasi-elastic intensity is also substantially weaker in Pr$_4$Ni$_3$O$_{10}$, although its sensitivity to sample-surface quality prevents a quantitative physical interpretation of this difference. In contrast to the Ni $L_3$-edge differences, the normalized O $K$-edge spectra in Fig.~\ref{fig:S5_RIXS_example}(c) exhibit nearly identical line shapes when measured at the same incident angle, resonant incident energy, and polarization. Such agreement indicates that the energy distribution of the O-derived charge-transfer excitations, and therefore the overall O $2p$ contribution to the hybridized Ni--O electronic states, remains similar in the two compounds.\vspace{-5mm}

\section{Low-energy O $K$-edge RIXS response}
\label{sec:OK_phonon}\vspace{-3mm}

\begin{figure}
    \centering
    \includegraphics[width=0.7\columnwidth]
    {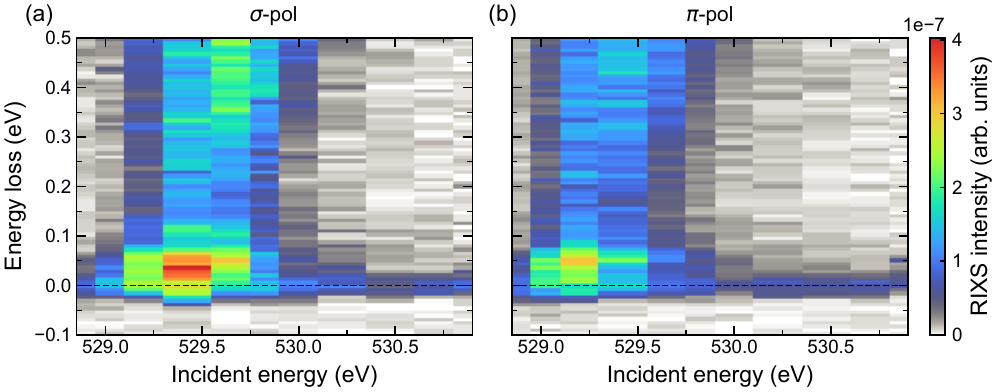}
    \caption{Low-energy O $K$-edge RIXS response of Pr$_4$Ni$_3$O$_{10}$ as a function of incident photon energy for (a) $\sigma$ and (b) $\pi$ polarization. The maps show enlarged views of the low-energy incident-energy-dependent RIXS response. A phonon excitation near 50~meV appears around the O $K$ pre-edge resonance and is more pronounced for $\sigma$ polarization.}\vspace{-3mm}
    \label{fig:S6_OK_Emap_zoom}
\end{figure}

O $K$-edge RIXS is directly sensitive to oxygen-derived states and readily resolves low-energy phonons. Previous measurements on RP nickelates identified phonon excitations in the 50--70~meV range and, in La$_4$Ni$_3$O$_{10}$, modes near 53 and 70~meV were associated with out-of-plane bond buckling and in-plane bond stretching, respectively~\cite{he2026layer,zhang2026distinct}. Figure~\ref{fig:S6_OK_Emap_zoom} shows enlarged low-energy O $K$-edge RIXS maps of Pr$_4$Ni$_3$O$_{10}$ for both incident polarizations. As we tune the incident energy through the O $K$- pre-edge resonance, we observe an excitation near 50~meV, consistent with a phonon mode, which is more pronounced for $\sigma$ than for $\pi$ polarization. Owing to the limited measurement time and modest counting statistics, however, we cannot assign this feature unambiguously to a specific phonon branch or quantify its polarization dependence.

\section{Momentum-Dependent RIXS Measurement Geometry}\vspace{-3mm}

\begin{figure}[h]
    \centering
    \includegraphics[width=0.8\textwidth]{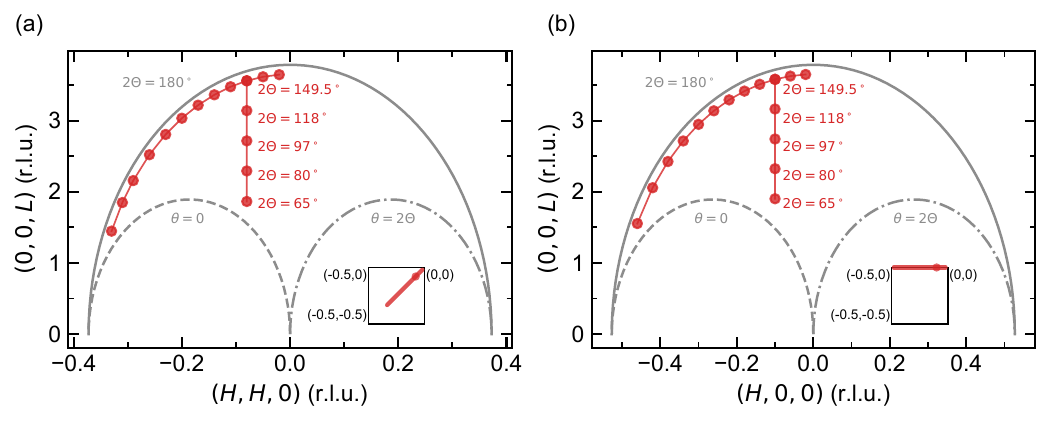}
    \caption{Accessible momentum-space regions and measured Ni $L_3$-edge RIXS points in the (a) $(H,H,L)$ and (b) $(H,0,L)$ scattering planes. Gray curves delimit the kinematically accessible region: the outer boundary corresponds to $2\Theta=180^\circ$, while the inner boundaries mark $\theta=0$ and $\theta=2\Theta$, where the incident or scattered beam becomes parallel to the sample surface. Red dots indicate the measured momentum points. Insets show the corresponding in-plane momentum trajectories, with red dots marking the fixed in-plane momenta used for the $L$-dependent scans.}
    \label{fig:S7_Qpoints}
\end{figure}

Momentum-dependent Ni $L_3$-edge RIXS measurements were performed in the $(H,H,L)$ and $(H,0,L)$ scattering planes. We rotated the sample azimuthal angle $\phi$ to switch between the two geometries and measure the in-plane momentum dependence along $[H,H]$ and $[H,0]$, respectively. Figure~\ref{fig:S7_Qpoints} summarizes the accessible momentum-space regions and measured points in both geometries. For the in-plane scans, we fixed the spectrometer scattering angle at $2\Theta=149.5^\circ$ and varied the sample angle $\theta$. This scans the in-plane momentum transfer along the selected high-symmetry direction, while the corresponding $L$ value is set by the scattering geometry. The measured points are shown by red dots along the $2\Theta=149.5^\circ$ curves in Fig.~\ref{fig:S7_Qpoints}. For the out-of-plane scans, we varied $\theta$ and $2\Theta$ simultaneously to change $L$ while keeping the in-plane momentum transfer fixed. Measurements were performed at $q_{\parallel}=(-0.08,-0.08)$ r.l.u. in the $(H,H,L)$ plane and $q_{\parallel}=(-0.10,0)$ r.l.u. in the $(H,0,L)$ plane. The corresponding points are shown by red dots at fixed in-plane momentum in Fig.~\ref{fig:S7_Qpoints}.\vspace{-3mm}

\section{Momentum Dependence of the Low-Energy RIXS Response}\vspace{-3mm}

\begin{figure}[h]
    \centering
    \includegraphics[width=0.9\textwidth]
    {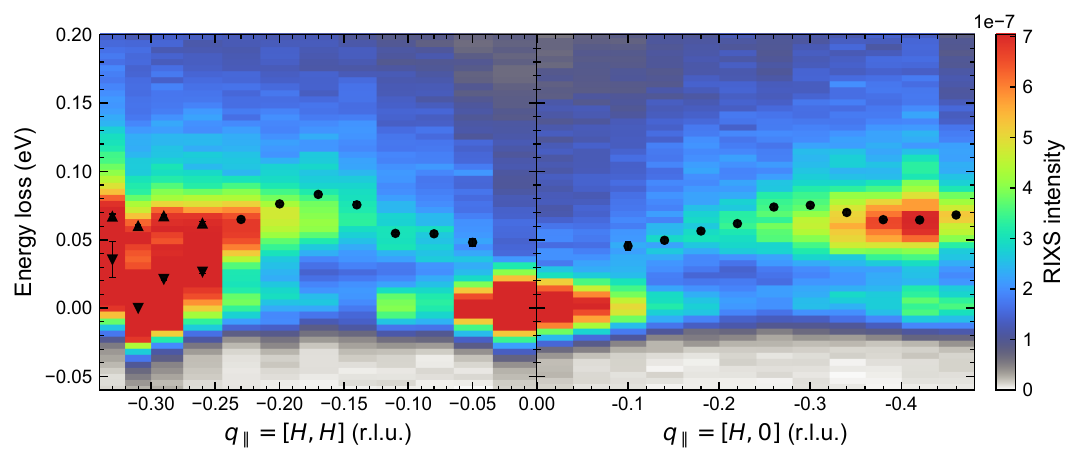}
    \caption{In-plane momentum dependence of the low-energy Ni $L_3$-edge RIXS response of Pr$_4$Ni$_3$O$_{10}$ along $[H,H]$ and $[H,0]$, measured with $\pi$-polarized incident x-rays at the magnon resonance energy. The maps retain the quasi-elastic response near zero energy loss. Black symbols mark the fitted low-energy magnon energies. Where two modes are resolved, both are shown.}\vspace{-3mm}
    \label{fig:S8_Qmap_HHL_H0L}
\end{figure}

\begin{figure}[h]
    \centering
    \includegraphics[width=0.75\textwidth]
    {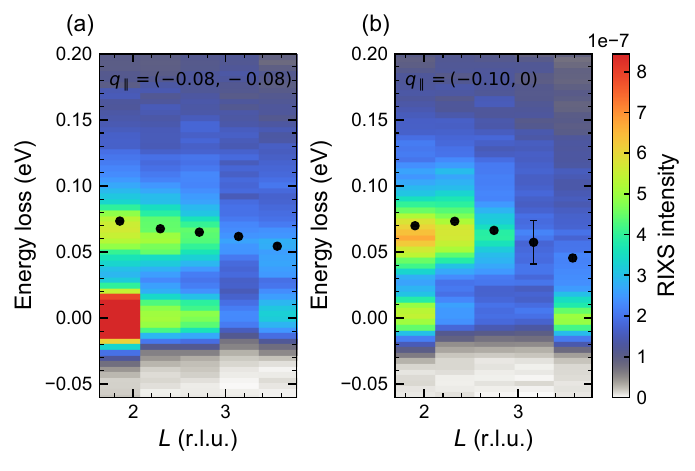}
    \caption{Out-of-plane momentum dependence of the low-energy Ni $L_3$-edge RIXS response of Pr$_4$Ni$_3$O$_{10}$ at fixed in-plane momentum transfer. (a) $(H,H,L)$ plane at $q_{\parallel}=(-0.08,-0.08)$ r.l.u. (b) $(H,0,L)$ plane at $q_{\parallel}=(-0.10,0)$ r.l.u. The spectra retain the quasi-elastic response. Black symbols mark the fitted low-energy magnon energies; vertical error bars indicate the corresponding fitting uncertainties.}\vspace{-3mm}
    \label{fig:S9_Ldeps}
\end{figure}

\begin{figure}[h]
    \centering
    \includegraphics[width=0.8\textwidth]
    {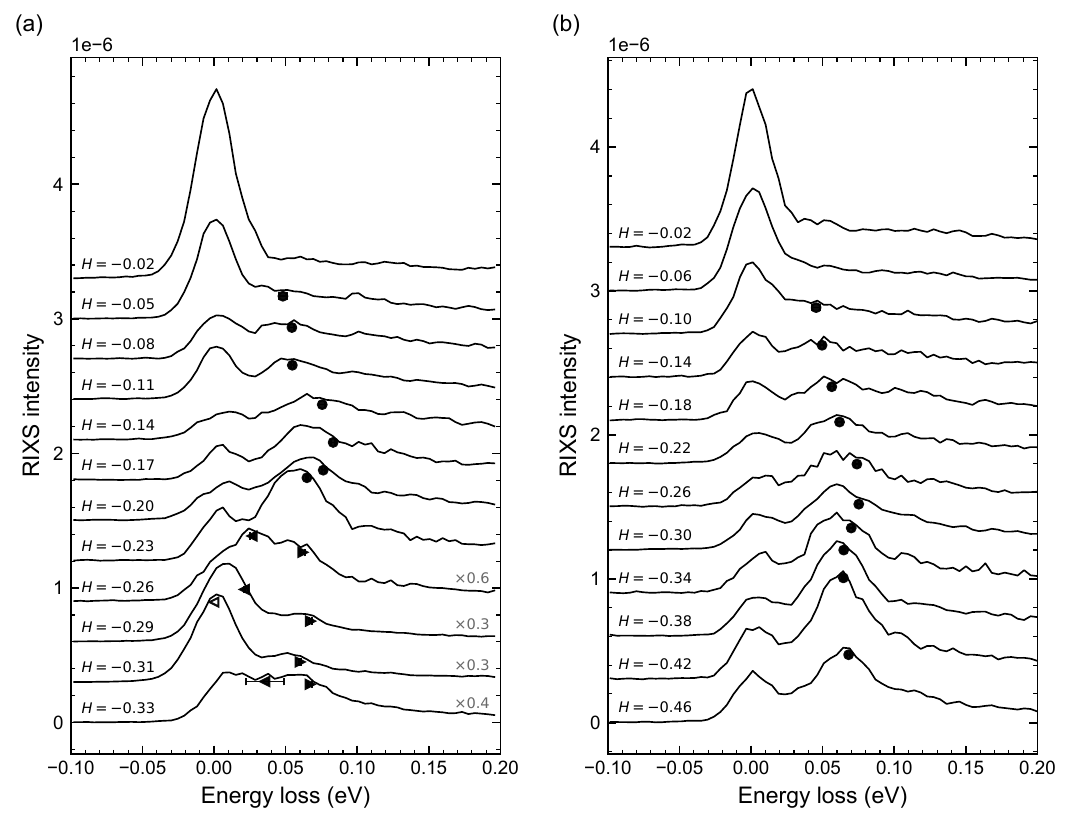}
    \caption{In-plane momentum-dependent Ni $L_3$-edge RIXS spectra along (a) $q_{\parallel}=(H,H)$ and (b) $q_{\parallel}=(H,0)$. Successive spectra are vertically offset for clarity. Black symbols mark the fitted low-energy magnon energies, with different symbol shapes distinguishing the two modes where both are resolved. Horizontal error bars indicate the corresponding fitting uncertainties. Selected spectra near the SDW momentum in panel (a) were multiplied by the indicated factors for clarity.}\vspace{-3mm}
    \label{fig:S10_Waterfall_HHL_H0L}
\end{figure}

\begin{figure}[h]
    \centering
    \includegraphics[width=0.75\textwidth]
    {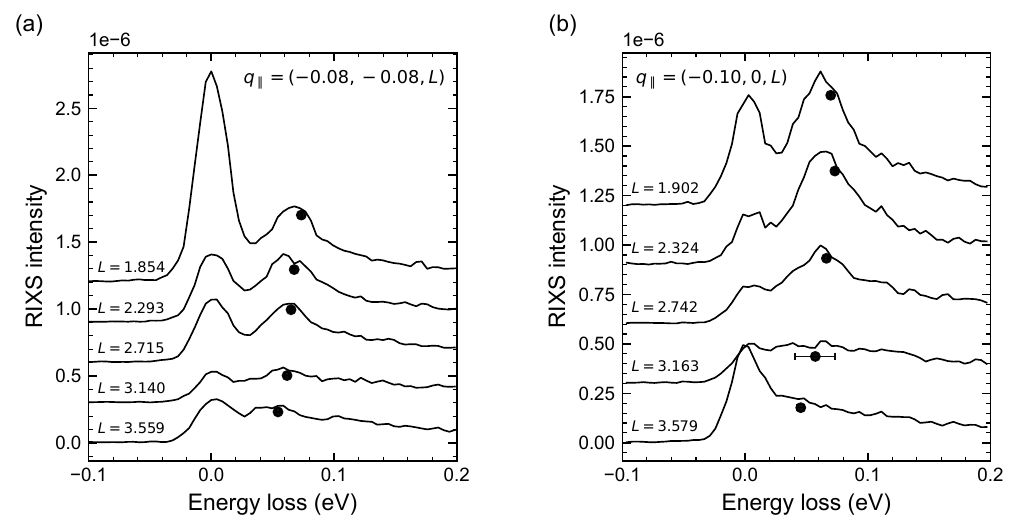}
    \caption{Out-of-plane momentum-dependent Ni $L_3$-edge RIXS spectra at (a) $q_{\parallel}=(-0.08,-0.08)$ r.l.u. in the $(H,H,L)$ plane and (b) $q_{\parallel}=(-0.10,0)$ r.l.u. in the $(H,0,L)$ plane. Successive spectra are vertically offset for clarity. Black symbols mark the fitted low-energy magnon energies. Horizontal error bars indicate the corresponding fitting uncertainties. Broad higher-energy magnetic contributions included in selected fits are not shown.}\vspace{-3mm}
    \label{fig:S11_Waterfall_Ldeps}
\end{figure}

Figures~\ref{fig:S8_Qmap_HHL_H0L}--\ref{fig:S11_Waterfall_Ldeps} show the momentum dependence of the low-energy Ni $L_3$-edge RIXS response measured with $\pi$-polarized incident x-rays at the magnon resonance energy. In contrast to the corresponding maps in the main text, these spectra retain the quasi-elastic response near zero energy loss. Figure~\ref{fig:S8_Qmap_HHL_H0L} shows instead the in-plane response along $[H,H]$ and $[H,0]$. In both directions, we observe a dispersive magnetic excitation with energies spanning approximately 20--85~meV. Along $[H,H]$, the mode softens toward the spin-density-wave (SDW) momentum near $H\simeq-0.31$ r.l.u., where spectral weight also accumulates near zero energy loss. Figure~\ref{fig:S9_Ldeps} shows the corresponding out-of-plane response at fixed in-plane momentum, with $q_{\parallel}=(-0.08,-0.08)$ r.l.u. in the $(H,H,L)$ plane and $q_{\parallel}=(-0.10,0)$ r.l.u. in the $(H,0,L)$ plane. The magnetic response varies with $L$ primarily through changes in line shape and spectral weight, while the fitted excitation energies vary more weakly, revealing a finite out-of-plane dependence of the low-energy response. The waterfall plots in Figs.~\ref{fig:S10_Waterfall_HHL_H0L} and \ref{fig:S11_Waterfall_Ldeps} provide complementary views of the same datasets and make the evolution of the quasi-elastic response and magnetic peaks directly visible. Some spectra also require a broad higher-energy magnetic component to the fits, which we attribute to a multimagnon response. This contribution is shown only in the fit decompositions in Figs.~\ref{fig:S14_HHL_fits}--\ref{fig:S17_H0L_Ldep_fits}. 

\FloatBarrier
\begin{figure}[h]
    \centering
    \includegraphics[width=0.75\textwidth]
    {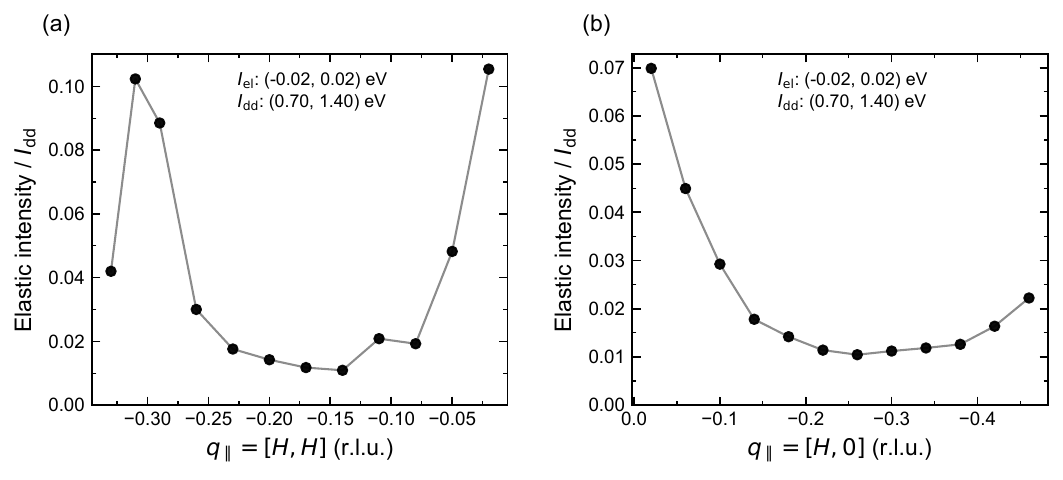}
    \caption{Momentum dependence of the integrated quasi-elastic RIXS intensity along (a) $q_{\parallel}=(H,H)$ and (b) $q_{\parallel}=(H,0)$. For each spectrum, we obtained $I_{\mathrm{el}}$ by integrating the measured intensity from $-0.02$ to $0.02$~eV and normalized it by the integrated $dd$-excitation intensity, $I_{dd}$, between $0.70$ and $1.40$~eV. The plotted values are obtained directly by numerical integration and do not rely on spectral fits. In panel (a), the enhancement near the SDW momentum reflects magnon softening and the quasi-static SDW response, while the strong intensity near $\mathbf{q}_{\parallel}=0$ is dominated by specular reflection. The weaker increase in panel (b) may include a contribution from diffuse surface scattering.}\vspace{-3mm}
    \label{fig:S12_ElasticIntensity}
\end{figure}

Figure~\ref{fig:S12_ElasticIntensity} shows the momentum dependence of the near-zero-energy intensity. For each spectrum, we obtained the quasi-elastic intensity, $I_{\mathrm{el}}$, by integrating the RIXS signal from $-0.02$ to $0.02$~eV and normalized it by the integrated $dd$-excitation intensity, $I_{\mathrm{dd}}$, between $0.70$ and $1.40$~eV, $I_{\mathrm{el}}/I_{dd}=\displaystyle\int_{-0.02\,\mathrm{eV}}^{0.02\,\mathrm{eV}} I(\omega)\,d\omega\bigg/\displaystyle\int_{0.70\,\mathrm{eV}}^{1.40\,\mathrm{eV}} I(\omega)\,d\omega$. This quantity is obtained directly by numerical integration and does not depend on the spectral fits. Along $[H,H]$, the pronounced enhancement near the SDW momentum reflects the transfer of magnetic spectral weight toward zero energy as the dispersive mode softens toward the quasi-static SDW response. The strong intensity near $q_{\parallel}=0$ arises from specular reflection. Along $[H,0]$, the weaker increase in near-zero-energy intensity toward larger momentum transfer may originate from diffuse surface scattering associated with imperfect surface quality under low-scattering angle conditions.

\vspace{-3mm}
\section{Fitting the Low-Energy RIXS Spectra}\vspace{-3mm}
\label{sec:QmapFit}

\begin{figure}[h]
    \centering
    \includegraphics[width=0.8\textwidth]
    {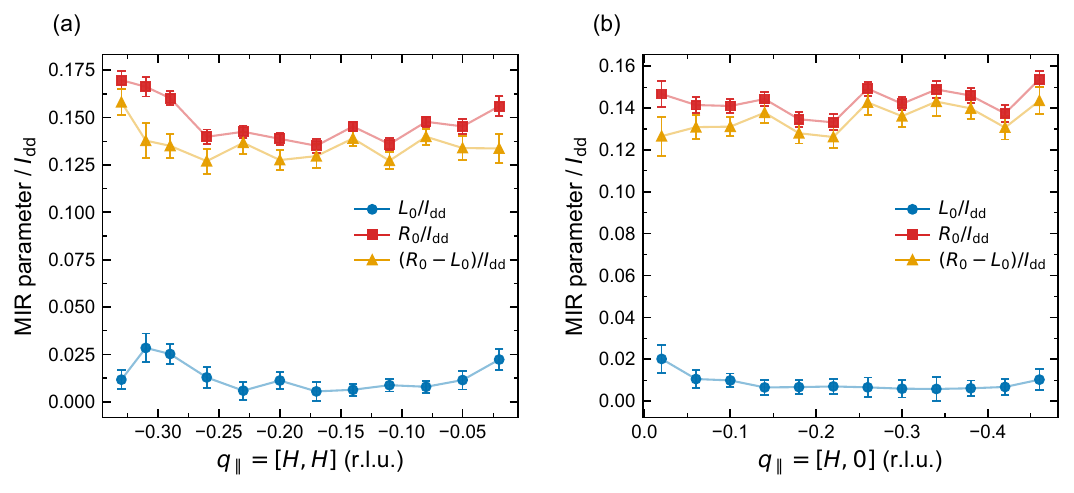}
    \caption{
    Momentum dependence of the fitted MIR-background parameters along (a) $q_{\parallel}=(H,H)$ and (b) $q_{\parallel}=(H,0)$. Blue circles, red squares, and yellow triangles show $L_0/I_{dd}$, $R_0/I_{dd}$, and $(R_0-L_0)/I_{dd}$, respectively. Here, $L_0$ and $R_0$ are the fitted background levels on the energy-gain and energy-loss sides of the quasi-elastic line, and $R_0-L_0$ is the magnitude of the step-like MIR contribution. The parameters are normalized by the integrated $dd$-excitation intensity $I_{dd}$ between $0.70$ and $1.40$~eV. Error bars represent fit uncertainties.
}
    \label{fig:S13_MIR_intensity}
\end{figure}

\begin{figure}[h]
    \centering
    \includegraphics[width=\textwidth]
    {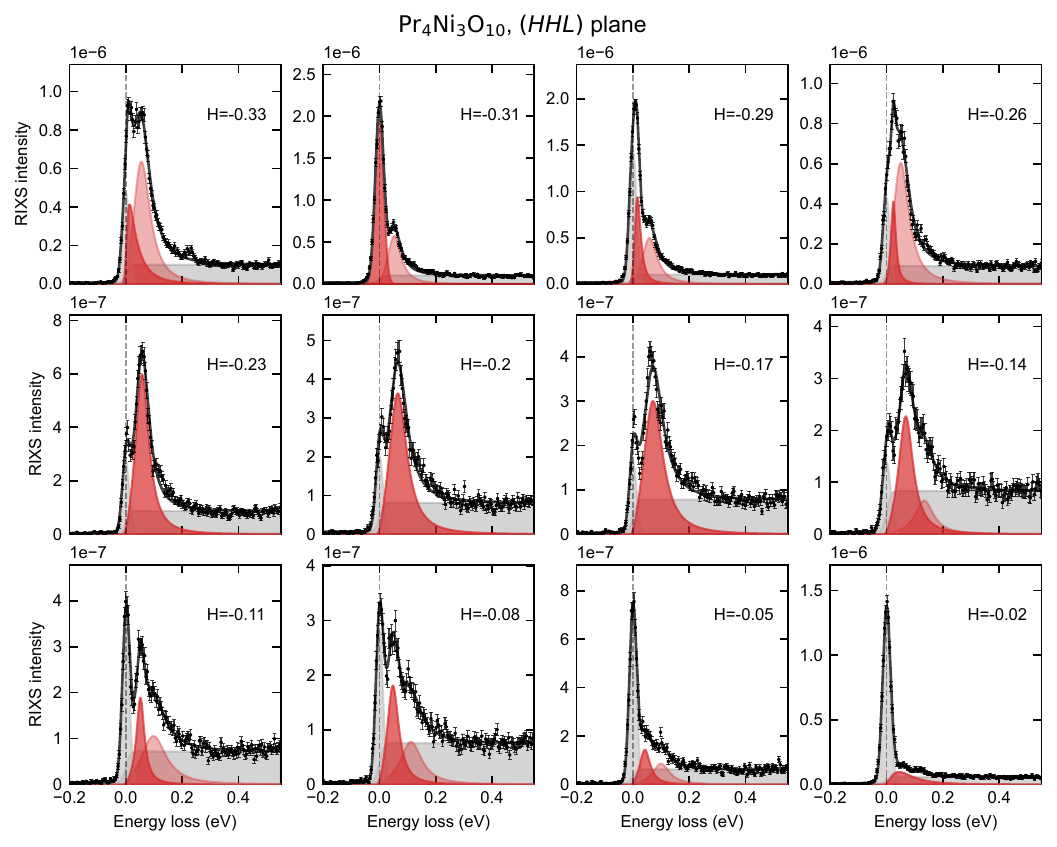}
    \caption{Fits to the in-plane momentum-dependent Ni $L_3$-edge RIXS spectra along $q_{\parallel}=(H,H)$. Black symbols represent the measured spectra and solid curves the total fits. Shaded components denote the Gaussian quasi-elastic line, step-like MIR background, and DHO contributions describing one or two low-energy magnon modes or a broad higher-energy multimagnon contribution, depending on momentum. Each energy-loss axis is aligned to the fitted quasi-elastic-line center, marked by the vertical dashed line at zero energy loss. Error bars represent statistical uncertainties.}
    \label{fig:S14_HHL_fits}
\end{figure}

\begin{figure}[h]
    \centering
    \includegraphics[width=\textwidth]
    {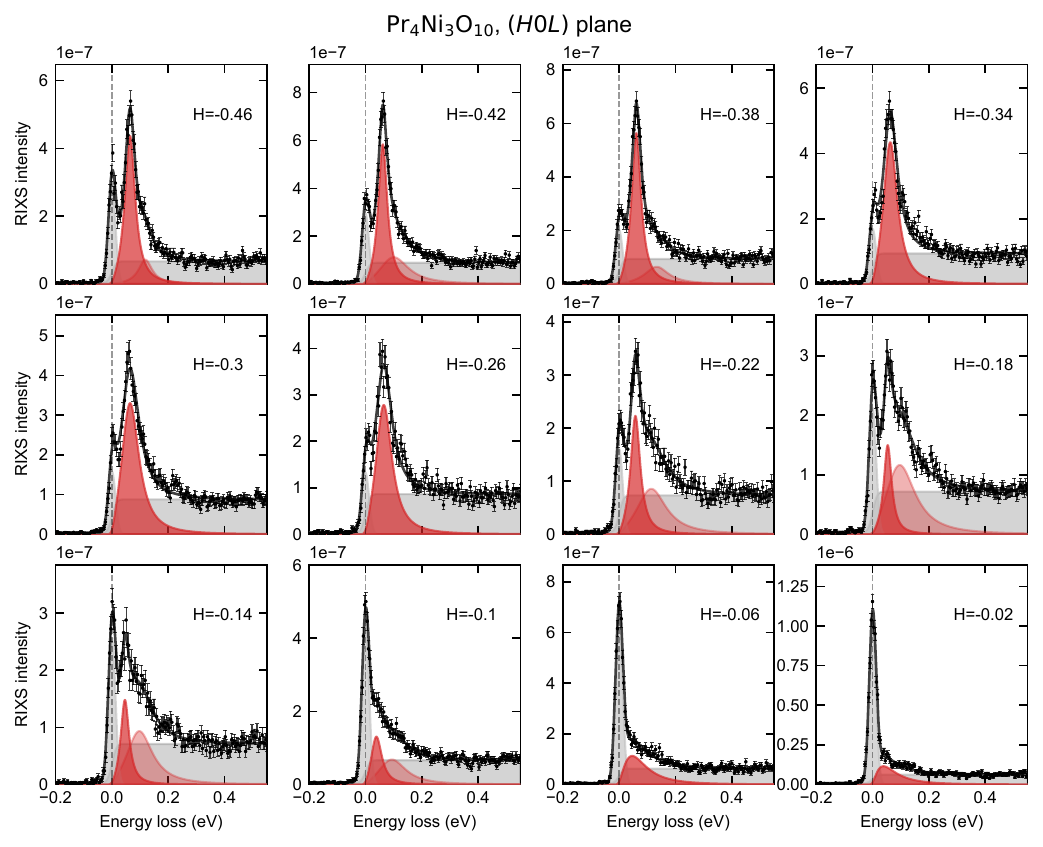}
    \caption{Fits to the in-plane momentum-dependent Ni $L_3$-edge RIXS spectra along $q_{\parallel}=(H,0)$. Black symbols represent the measured spectra and solid curves the total fits. Shaded components denote the Gaussian quasi-elastic line, step-like MIR background, and DHO contributions describing one or two low-energy magnon modes or a broad higher-energy multimagnon contribution, depending on momentum. Each energy-loss axis is aligned to the fitted quasi-elastic-line center, marked by the vertical dashed line at zero energy loss. Error bars represent statistical uncertainties.}
    \label{fig:S15_H0L_fits}
\end{figure}

\begin{figure}
    \centering
    \includegraphics[width=0.65\textwidth]
    {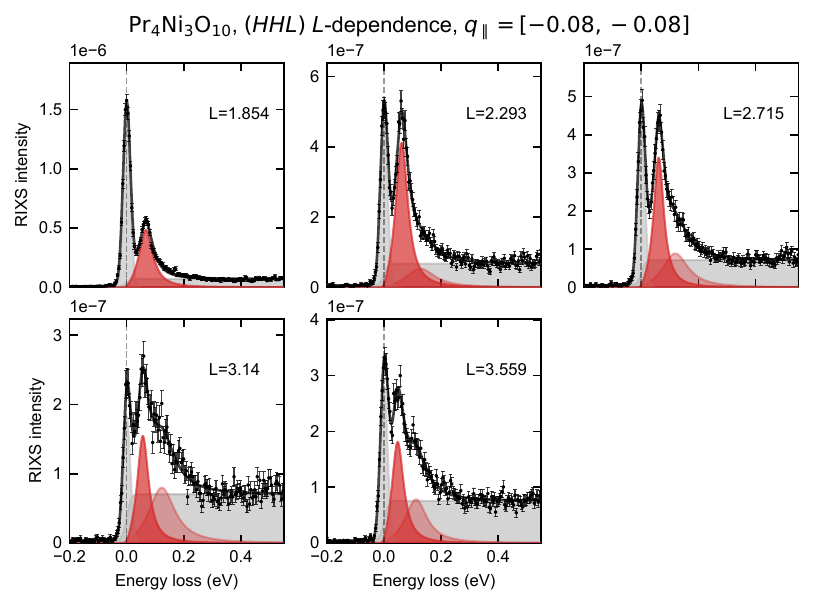}
    \caption{Fits to the out-of-plane momentum-dependent Ni $L_3$-edge RIXS spectra in the $(H,H,L)$ plane at fixed $q_{\parallel}=(-0.08,-0.08)$ r.l.u. Black symbols represent the measured spectra and solid curves the total fits. Shaded components denote the Gaussian quasi-elastic line, step-like MIR background, and DHO contributions describing one or two low-energy magnon modes or a broad higher-energy multimagnon contribution, depending on momentum. Each energy-loss axis is aligned to the fitted quasi-elastic-line center, marked by the vertical dashed line at zero energy loss. Error bars represent statistical uncertainties.}
    \label{fig:S16_HHL_Ldep_fits}
\end{figure}

\begin{figure}
    \centering
    \includegraphics[width=0.65\textwidth]
    {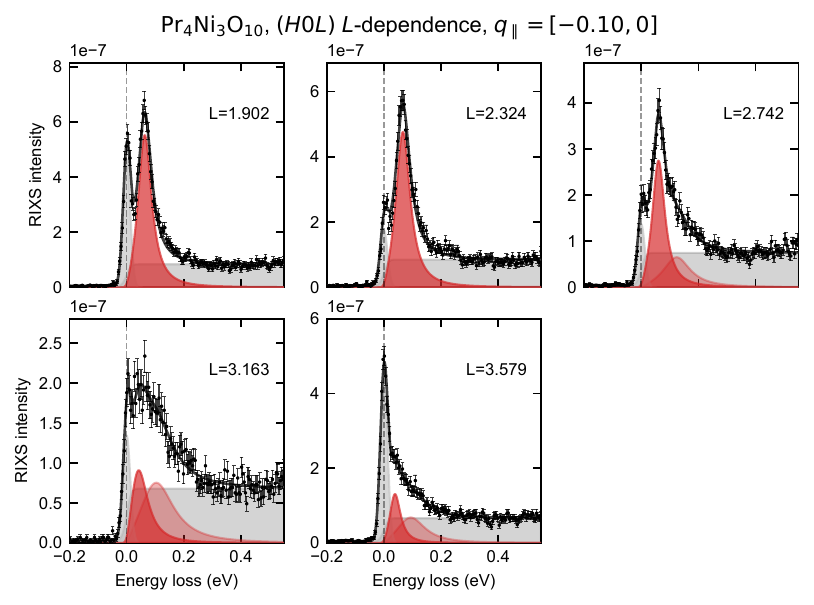}
    \caption{Fits to the out-of-plane momentum-dependent Ni $L_3$-edge RIXS spectra in the $(H,0,L)$ plane at fixed $q_{\parallel}=(-0.10,0)$ r.l.u. Black symbols represent the measured spectra and solid curves the total fits. Shaded components denote the Gaussian quasi-elastic line, step-like MIR background, and DHO contributions describing one or two low-energy magnon modes or a broad higher-energy multimagnon contribution, depending on momentum. Each energy-loss axis is aligned to the fitted quasi-elastic-line center, marked by the vertical dashed line at zero energy loss. Error bars represent statistical uncertainties.}
    \label{fig:S17_H0L_Ldep_fits}
\end{figure}

We fitted the momentum-dependent RIXS spectra over the energy-loss range from $-0.20$ to $0.55$~eV. For each spectrum, we determined the elastic-line position from the fit and shifted the energy-loss axis so that its center lies at zero energy. All spectra, fit components, and extracted excitation energies in Figs.~\ref{fig:S8_Qmap_HHL_H0L}--\ref{fig:S17_H0L_Ldep_fits} are shown on this aligned energy-loss scale.

The fitting model comprises a Gaussian quasi-elastic line, one or two damped harmonic oscillator (DHO) contributions, and a smooth error-function step describing the mid-infrared (MIR) background:

\begin{equation}
I(\omega)
=
I_{\mathrm{el}}(\omega)
+
I_{\mathrm{MIR}}(\omega)
+
\sum_{j=1}^{N_{\mathrm{DHO}}}
I_{\mathrm{DHO},j}(\omega),
\qquad
N_{\mathrm{DHO}}=1\ \mathrm{or}\ 2.
\label{eq:total_rixs_fit}
\end{equation}

Here, $\omega$ denotes energy loss after aligning the fitted quasi-elastic-line center to zero. The quasi-elastic contribution was modeled as a Gaussian,

\begin{equation}
I_{\mathrm{el}}(\omega)
=
A_{\mathrm{el}}
\exp\left(
-\frac{\omega^2}{2\sigma_{\mathrm{el}}^2}
\right),
\label{eq:elastic_fit}
\end{equation}

where $A_{\mathrm{el}}$ and $\sigma_{\mathrm{el}}$ denote the fitted quasi-elastic-line amplitude and width, respectively. We modeled each magnetic contribution with a DHO response,

\begin{equation}
I_{\mathrm{DHO},j}(\omega)
=
A_j
\frac{\sigma_j\omega}
{\left(\omega^2-\omega_{0,j}^2\right)^2
+4\sigma_j^2\omega^2}
\frac{1}{1-\exp(-\omega/k_{\mathrm{B}}T)} ,
\label{eq:dho_fit}
\end{equation}

where $A_j$ sets the intensity, $\omega_{0,j}$ is the undamped mode energy, and $\sigma_j$ controls the damping. In conventional notation, $\Gamma_j=2\sigma_j$. We evaluated the detailed-balance factor at the measurement temperature, $T=30$~K. The number and interpretation of the DHO components depend on the spectral line shape. We use one or two DHO terms when one or two low-energy magnon modes are resolved. When no low-energy magnon is clearly resolved, we retain a weak, broad DHO to represent the higher-energy multimagnon response. We add the same component when residual magnetic spectral weight remains above a resolved magnon mode. The symbols in Figs.~\ref{fig:S8_Qmap_HHL_H0L}--\ref{fig:S11_Waterfall_Ldeps} show only the fitted low-energy magnon modes, including both energies where two modes are resolved. Broad DHO components describing the higher-energy multimagnon response are omitted from the magnon dispersion and shown only in the fit decompositions in Figs.~\ref{fig:S14_HHL_fits}--\ref{fig:S17_H0L_Ldep_fits}. We modeled the MIR background as a smooth step centered at zero energy loss,

\begin{equation}
I_{\mathrm{MIR}}(\omega)
=
L_0+(R_0-L_0)S(\omega),
\qquad
S(\omega)
=
\frac{1}{2}
\left[
1+
\operatorname{erf}
\left(
\frac{\omega}{\sqrt{2}\sigma_{\mathrm{el}}}
\right)
\right].
\label{eq:mir_step}
\end{equation}

Here, $L_0$ and $R_0$ are the background levels on the energy-gain and energy-loss sides of the quasi-elastic line, respectively, and $R_0-L_0$ gives the magnitude of the step-like MIR contribution. We set the smoothing width of the error-function step equal to the fitted quasi-elastic-line width, $\sigma_{\mathrm{el}}$, thereby tying the MIR onset to the fitted zero-energy position without introducing an additional width parameter.

Figure~\ref{fig:S13_MIR_intensity} shows the momentum dependence of the fitted MIR-background parameters for the in-plane scans. To account for variations in the overall RIXS intensity, we normalized $L_0$, $R_0$, and $R_0-L_0$ by the integrated $dd$-excitation intensity, $I_{dd}$, between $0.70$ and $1.40$~eV. The finite $(R_0-L_0)/I_{dd}$ across the measured momentum range is consistent with a broad metallic MIR background, whose weak momentum dependence distinguishes it from the dispersive low-energy magnetic response. Figures~\ref{fig:S14_HHL_fits}--\ref{fig:S17_H0L_Ldep_fits} show the complete spectral fits. Figures~\ref{fig:S14_HHL_fits} and \ref{fig:S15_H0L_fits} correspond to the in-plane scans along $[H,H,L]$ and $[H,0,L]$, respectively, while Figs.~\ref{fig:S16_HHL_Ldep_fits} and \ref{fig:S17_H0L_Ldep_fits} show the out-of-plane scans at fixed $q_{\parallel}=(-0.08,-0.08)$ r.l.u. and $q_{\parallel}=(-0.10,0)$ r.l.u. Each decomposition includes the quasi-elastic line, MIR background, and required DHO components, which represent one or two low-energy magnon modes or a broad higher-energy multimagnon contribution, depending on momentum.

\FloatBarrier

\section{Linear-Spin-Wave Modeling of the Magnetic Excitations}\label{sec:LSWT}

To describe the momentum-dependent magnetic response, we used linear spin-wave theory (LSWT) for an effective Heisenberg model of the spin-density-wave state. Although Pr$_4$Ni$_3$O$_{10}$ is metallic and its magnetic order may have a substantial itinerant component, this model provides a minimal phenomenological description of the collective excitation spectrum. Similar Heisenberg descriptions have been used for magnetic excitations in both bilayer and trilayer Ruddlesden--Popper nickelates~\cite{tenhuisen2025magnetic,chen2024electronic,chan2026collective,chen2026dissecting}.

\subsection{Effective exchange model}

The magnetic unit comprises the two outer NiO$_2$ layers of a trilayer block. We assign no ordered moment to the middle NiO$_2$ layer, consistent with diffraction measurements placing it near a node of the SDW modulation~\cite{samarakoon2023bootstrapped}. The two outer layers carry oppositely phased ordered moments coupled by an effective interlayer exchange interaction. The Hamiltonian is\vspace{-3mm}

\begin{equation}
\mathcal{H}
=
J_1\sum_{\ell=1,2}\sum_{\langle i,j\rangle}\mathbf{S}_{i,\ell}\cdot\mathbf{S}_{j,\ell}
+
J_2\sum_{\ell=1,2}\sum_{\langle\!\langle i,j\rangle\!\rangle}\mathbf{S}_{i,\ell}\cdot\mathbf{S}_{j,\ell}
+
J_3\sum_{\ell=1,2}\sum_{\langle\!\langle\!\langle i,j\rangle\!\rangle\!\rangle}\mathbf{S}_{i,\ell}\cdot\mathbf{S}_{j,\ell}
+
J_z\sum_i\mathbf{S}_{i,1}\cdot\mathbf{S}_{i,2},
\label{eq:lswt_hamiltonian}
\end{equation}

where $\ell=1,2$ labels the two outer NiO$_2$ layers. Within each layer, $J_1$, $J_2$, and $J_3$ couple first-, second-, and third-nearest-neighbor Ni sites. The parameter $J_z$ couples vertically aligned sites in the two outer layers. Positive values denote antiferromagnetic exchange. The effective outer-layer coupling $J_z$ should not be interpreted as a direct exchange through the nominally nonmagnetic middle Ni site. Instead, it represents the net interaction between the magnetic outer-layer degrees of freedom after integrating out the middle-layer electronic states. A sizable $J_z$ is physically plausible because the partially occupied Ni $3d_{3z^2-r^2}$ orbitals hybridize strongly along the $c$ axis through apical O $2p_z$ states. Hence, the resulting interlayer hopping can generate an appreciable magnetic interaction even when the static ordered moment of the middle layer is strongly suppressed. This mechanism is closely related to the strong interlayer coupling discussed in other Ruddlesden--Popper nickelates~\cite{chen2024electronic,chen2026dissecting}. 

We use an effective spin length $S=1$, so the reported exchange parameters correspond to the energy scales $SJ_i$. Because Pr$_4$Ni$_3$O$_{10}$ is metallic, these parameters provide a phenomenological description of the collective SDW excitations rather than purely localized-spin superexchange constants. Within LSWT, we calculated the trace of the dynamical spin structure factor,

\begin{equation}
S_{\mathrm{tr}}(\mathbf{q},\omega)
=
\sum_{\alpha=x,y,z}
\frac{1}{2\pi N}
\sum_{i,j}
\int dt\,
e^{i[\omega t-\mathbf{q}\cdot(\mathbf{r}_i-\mathbf{r}_j)]}
\left\langle
S_i^\alpha(t)S_j^\alpha(0)
\right\rangle .
\label{eq:ssf_trace}
\end{equation}

We performed the LSWT calculations using the \textsc{Sunny} package~\cite{dahlbom2025sunny}. For each set of exchange parameters, we minimized the classical spiral ground state starting from a propagation vector near $(-0.31,-0.31,0)$ r.l.u. and evaluated the dynamical spin structure factor about the optimized state. This quantity contains the LSWT mode energies and momentum-dependent spectral weights, including interference between the two outer layers. We use it as an effective description of the magnetic contribution to the Ni $L_3$-edge RIXS response without explicitly calculating the full resonant scattering cross section.

\vspace{-3mm}
\subsection{Structural twin averaging}
\label{sec:lswt_domains}\vspace{-3mm}

\begin{figure}[h]
    \centering
    \includegraphics[width=0.75\textwidth]
    {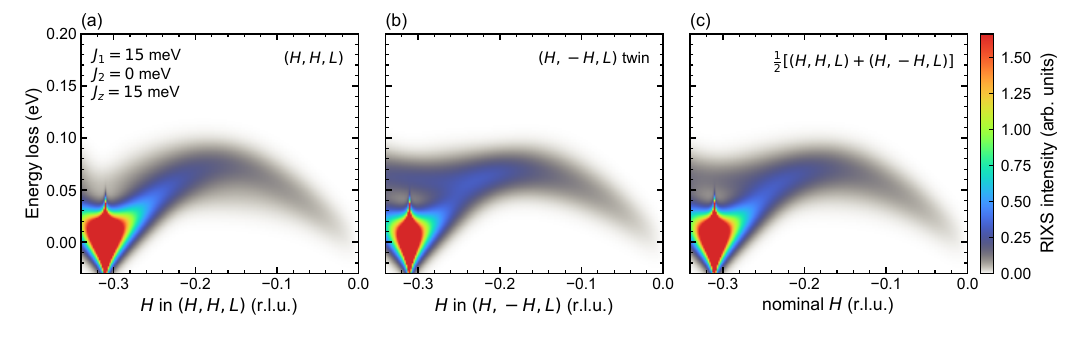}
    \caption{Effect of structural-twin averaging on the calculated LSWT response for $J_1=15$~meV, $J_2=0$, $J_3=10.189$~meV, and $J_z=15$~meV. (a) Response along $(H,H,L)$ in one structural twin. (b) Response along the corresponding $(H,-H,L)$ trajectory in the orthogonal twin. (c) Equal-weight incoherent average of the two twin responses according to Eq.~\eqref{eq:domain_average}. All panels use the same intensity scale.}\vspace{-7mm}
    \label{fig:S18_domain_average}
\end{figure}

The Pr$_4$Ni$_3$O$_{10}$ crystal contains two orthogonal structural twins related by an approximately $90^\circ$ in-plane rotation. In the pseudo-tetragonal notation used here, a nominal $(H,H,L)$ RIXS scan probes $\mathbf{q}=(H,H,L)$ in one twin and $\mathbf{q}=(H,-H,L)$ in the other. Because scattering from spatially separated twins adds incoherently, we average the calculated responses of the two domains. Assuming equal populations, the averaged in-plane intensity is
\begin{equation}
I_{\mathrm{avg}}(H,L,\omega)=\frac{1}{2}
\left[
I(H,H,L,\omega)+I(H,-H,L,\omega)
\right].
\label{eq:domain_average}
\end{equation}
Figure~\ref{fig:S18_domain_average} illustrates the effect of structural-twin averaging for a representative parameter set. Because the same experimental momentum trajectory samples different directions relative to the SDW propagation vector in the two orthogonal twins, their contributions to the measured dispersion differ. Along $(H,-H,L)$, a distinct higher-energy branch gains substantial spectral weight near 60~meV as $H$ approaches $-0.31$ r.l.u. The corresponding branch is also present along $(H,H,L)$ but carries much weaker weight in this region. The observed second magnetic branch near the SDW wave vector thus requires contributions from both twins. Unless otherwise stated, all subsequent calculations along the nominal $[H,H]$ direction are averaged over the two twins according to Eq.~\eqref{eq:domain_average}. We apply the same averaging to the $L$-dependent LSWT calculation at fixed $q_{\parallel}=(-0.08,-0.08)$ r.l.u. in the $(H,H,L)$ plane.

\subsection{Constraint imposed by the SDW wave vector}\vspace{-3mm}

The experimentally observed in-plane SDW propagation vector from neutron diffraction experiments is $\mathbf{q}_{\mathrm{SDW}}=(-\delta,-\delta,0)$, with $\delta=0.31$ ~\cite{samarakoon2023bootstrapped}. For a single-$\mathbf{q}$ magnetic modulation with in-plane propagation vector $\mathbf{q}=(h,k,0)$, the classical exchange energy is determined by the Fourier transform of the in-plane exchange interactions
\begin{equation}
J(h,k)=2J_1\left[\cos(2\pi h)+\cos(2\pi k)\right]+4J_2\cos(2\pi h)\cos(2\pi k)+2J_3\left[\cos(4\pi h)+\cos(4\pi k)\right].
\label{eq:Jq_general}
\end{equation}
Along the diagonal direction $h=k=\delta$, this becomes
\begin{equation}
J(\delta,\delta)=4J_1\cos(2\pi\delta)+4J_2\cos^2(2\pi\delta)+4J_3\cos(4\pi\delta).
\label{eq:Jq_diagonal}
\end{equation}
The ordering wave vector is obtained by minimizing this expression with respect to $\delta$. For an incommensurate solution with $\sin(2\pi\delta)\neq0$, the stationarity condition gives
\begin{equation}
J_1+2J_2\cos(2\pi\delta)+4J_3\cos(2\pi\delta)=0.
\label{eq:sdw_stationary}
\end{equation}
Consequently, $J_3=-\frac{J_1}{4\cos(2\pi\delta)}-\frac{J_2}{2}$, which for $\delta=0.31$ gives $J_3=0.67912\,J_1-0.5\,J_2$. We imposed this constraint throughout the fits to keep the observed SDW wave vector fixed. The independent fit parameters were therefore $J_1$, $J_2$, and $J_z$, while $J_3$ was recalculated from $J_1$ and $J_2$ at each iteration. This procedure prevents the fit from shifting the magnetic ordering vector simply to improve a local portion of the excitation spectrum.

\vspace{-3mm}
\subsection{Dependence of the calculated response on the exchange parameters}\vspace{-3mm}

\begin{figure}[h]
    \centering
    \includegraphics[width=0.9\textwidth]
    {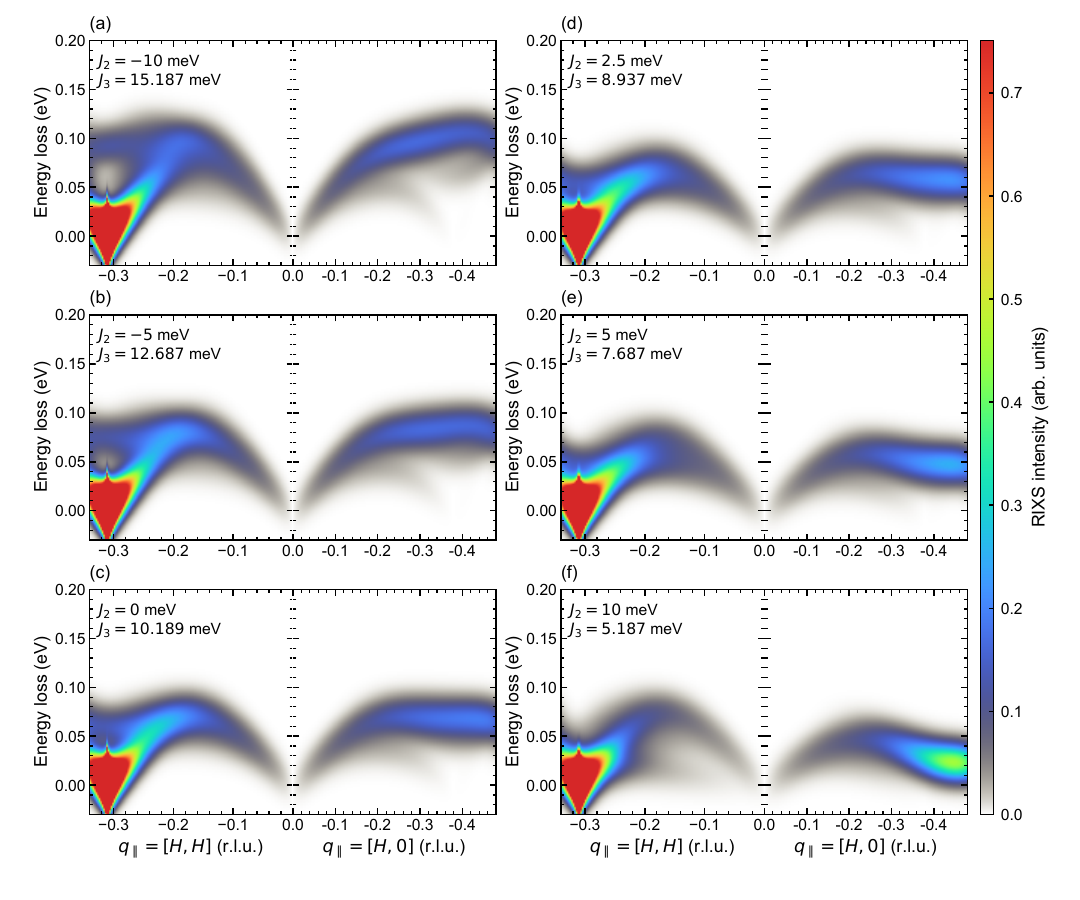}
    \caption{Dependence of the calculated in-plane LSWT response on $J_2$ and $J_3$. The calculations use $J_1=15$~meV and $J_z=15$~meV. For each parameter set, the left and right portions show the calculated response along $q_{\parallel}=[H,H]$ and $q_{\parallel}=[H,0]$, respectively. The values of $J_2$ and $J_3$ are constrained to preserve the experimental SDW wave vector $\delta=0.31$. The parameter sets are
    (a) $J_2=-10$~meV and $J_3=15.187$~meV,
    (b) $J_2=-5$~meV and $J_3=12.687$~meV, 
    (c) $J_2=0$ and $J_3=10.189$~meV,
    (d) $J_2=2.5$~meV and $J_3=8.937$~meV,
    (e) $J_2=5$~meV and $J_3=7.687$~meV, and
    (f) $J_2=10$~meV and $J_3=5.187$~meV.
Increasing $J_2$ while decreasing $J_3$ shifts the higher-energy branch near the SDW wave vector and the dominant $[H,0]$ response toward lower energy.
    }
    \label{fig:S19_LSWT_J2_J3}
\end{figure}

\begin{figure}[h]
    \centering
    \includegraphics[width=0.9\textwidth]
    {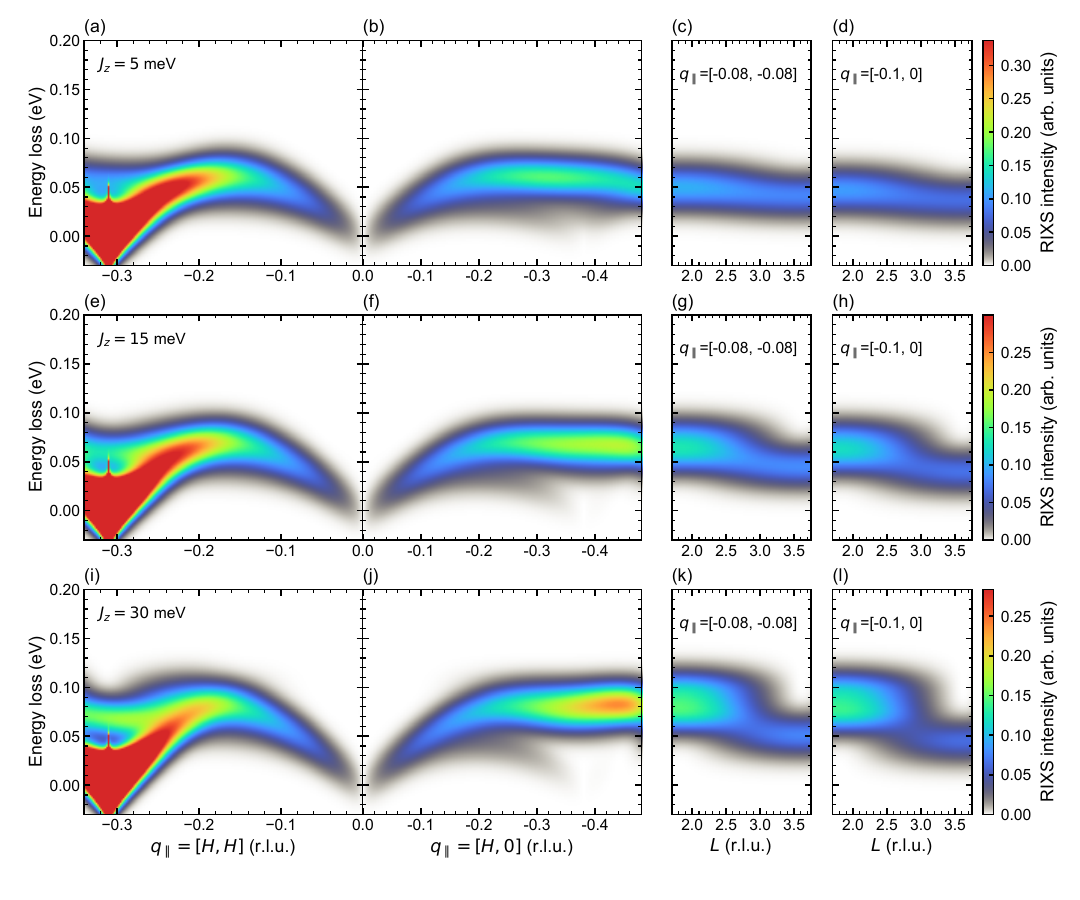}
    \caption{
    Dependence of the calculated LSWT response on the effective outer-layer exchange $J_z$. The calculations use $J_1=15$~meV, $J_2=0$, and $J_3=10.189$~meV, with $J_z=5$~meV in panels (a)--(d), $J_z=15$~meV in panels (e)--(h), and $J_z=30$~meV in panels (i)--(l). Within each row, the panels show the calculated response along $q_{\parallel}=[H,H]$, along $q_{\parallel}=[H,0]$, as a function of $L$ at fixed $q_{\parallel}=(-0.08,-0.08)$, and as a function of $L$ at fixed $q_{\parallel}=(-0.10,0)$, respectively. Increasing $J_z$ raises the dominant in-plane magnetic response and redistributes the $L$-dependent spectral weight toward higher-energy branches near $L\simeq2$. The response near $L\simeq3$--$4$ remains concentrated predominantly on the lower-energy branches.
    }
    \label{fig:S20_LSWT_Jz}
\end{figure}

To clarify how the different exchange interactions constrain the calculated spectrum, we separately examined the effects of the in-plane parameters $J_2$ and $J_3$ and of the interlayer parameter $J_z$. The results are summarized in Figs.~\ref{fig:S19_LSWT_J2_J3} and \ref{fig:S20_LSWT_Jz}.

Figure~\ref{fig:S19_LSWT_J2_J3} compares the calculated in-plane response for six combinations of $J_2$ and $J_3$. We fix $J_1=15$~meV and $J_z=15$~meV and vary $J_2$ and $J_3$ subject to the constraint imposed by the experimental SDW wave vector. All parameter sets reproduce the same ordering wave vector, while yielding substantially different excitation spectra. Along $[H,H]$, the overall bandwidth and softening toward the SDW wave vector remain similar across the constrained parameter sets. By contrast, the higher-energy branch near the SDW wave vector varies appreciably with $J_2$ and $J_3$. Increasing $J_2$ and decreasing $J_3$ shifts this branch to lower energy, following the same general trend as the dominant response along $[H,0]$, particularly at large $|H|$. The SDW wave vector constrains $J_1$, $J_2$, and $J_3$, while the measured excitation spectrum further refines their individual values. Figure~\ref{fig:S20_LSWT_Jz} illustrates the influence of $J_z$. Here, we fix $J_1=15$~meV, $J_2=0$, and $J_3=10.189$~meV and vary $J_z$ from 5 to 30~meV. Increasing $J_z$ raises the dominant magnetic response along both in-plane momentum directions. The effect is especially clear along $[H,0]$, where the main spectral feature shifts from approximately 50--60~meV for $J_z=5$~meV to approximately 80--90~meV for $J_z=30$~meV. Along $[H,H]$, the upper part of the magnetic response shifts upward similarly, while the soft mode at the SDW wave vector remains fixed by the imposed in-plane ordering condition. The $L$-dependent calculations show that $J_z$ primarily redistributes spectral weight among several spin-wave branches, whose eigenenergies remain only weakly dependent on $L$. Near $L\simeq2$, increasing $J_z$ transfers spectral weight from lower-energy modes to a higher-energy branch, which becomes particularly prominent for $J_z=30$~meV. Near $L\simeq3$--$4$, the dominant weight remains concentrated on a lower-energy branch and changes much less with $J_z$. Because the dynamical structure factors depend strongly on both $L$ and $J_z$, the dominant feature in the calculated maps can switch between branches even though the underlying mode energies remain nearly flat along $L$. This weak $L$ dispersion reflects the absence of intertrilayer exchange along $c$; the observed $L$ dependence instead arises mainly from interference between the two outer-layer scattering amplitudes, which redistributes spectral weight among the branches.

\vspace{-3mm}
\subsection{LSWT Fitting and Extracted Exchange Parameters}\vspace{-3mm}
We evaluated the response at the measured $\mathbf{q}=(H,K,L)$ points, including the variation of $L$ along the fixed-scattering-angle in-plane scans. The simultaneous fit included the in-plane scans along $[H,H]$ and $[H,0]$ and the $L$-dependent scans at fixed $q_{\parallel}=(-0.08,-0.08)$ r.l.u. and $(-0.10,0)$ r.l.u. Before comparing with the experiment, we convolved the calculated spectra with a 30~meV FWHM Gaussian, chosen to represent the typical quasi-elastic-line width obtained from the RIXS fits. We fitted the quasi-elastic-line-subtracted RIXS maps over the energy-loss range from 0 to 200~meV, after linearly scaling the calculated LSWT intensity to the experimental intensity. The exchange parameters were determined by a simultaneous least-squares fit to the two in-plane and two $L$-dependent momentum cuts. For the $(H,H,L)$ datasets, we averaged the calculated response over the two structural twins as described in Sec.~\ref{sec:lswt_domains}. The simultaneous fit gives
\begin{equation}
J_1=15.7\pm0.3~\mathrm{meV}\quad
J_2=1.7\pm0.3~\mathrm{meV}\quad
J_3=9.8\pm0.1~\mathrm{meV}\quad
J_z=15.8\pm0.6~\mathrm{meV}.
\end{equation}
The quoted uncertainties are one-standard-deviation estimates from the local least-squares fit and do not include systematic uncertainties associated with the assumed magnetic model or data treatment. The fitted interlayer interaction in Pr$_4$Ni$_3$O$_{10}$ is substantially smaller than the dominant interlayer exchange reported for bilayer La$_3$Ni$_2$O$_7$, where $J_z$ exceeds the in-plane couplings and strongly controls the magnetic response~\cite{chen2024electronic}. Instead, it is comparable to the interlayer scales reported for La$_4$Ni$_3$O$_{10}$, including $SJ_{\perp}\simeq20$~meV and $J_z\simeq22$~meV~\cite{chan2026collective,chen2026dissecting}. The remaining difference may reflect the different momentum coverage and fitting constraints, particularly our inclusion of explicit $L$-dependent measurements. In Pr$_4$Ni$_3$O$_{10}$, $J_1$, $J_3$, and $J_z$ all fall in the approximately 10--16~meV range, so no single interaction dominates the magnetic spectrum. The comparable in-plane and interlayer couplings indicate substantial contributions from both the Ni $3d_{x^2-y^2}$--O $2p_{x,y}$ network and the Ni $3d_{3z^2-r^2}$--O $2p_z$ channel. This interaction hierarchy distinguishes the trilayer compound from La$_3$Ni$_2$O$_7$ and points to a more balanced multiorbital magnetic network.

\vspace{-3mm}
\section{Entangled-dimer spin-wave calculations}\vspace{-3mm}
\label{sec:dimer}

Conventional LSWT treats the two outer-layer moments as separate semiclassical spins and expands about their classically ordered state. The interlayer exchange $J_z$ couples the moments and their spin-wave fluctuations, but the reference state does not include local entanglement within an outer-layer pair. To assess the possible role of such correlations, we also considered an entangled-dimer formulation in which the two outer-layer spins within each trilayer are treated as a single quantum unit~\cite{ganesh2011neel,lohofer2015dynamical}. For the $3\times3$ and $13\times13$ calculations below, the dimer model retains the same exchange network as conventional LSWT, with in-plane couplings $J_1$, $J_2$, and $J_3$ and an effective outer-layer coupling $J_z$. The distinction between the two calculations lies in the local basis. Conventional LSWT uses a product of individual spin states, whereas the dimer calculation allows each outer-layer pair to form an entangled local state. The corresponding local wave function can be written as $\left|\psi_{\mathrm{dimer}}\right\rangle=\sum_{m_1,m_2}c_{m_1m_2}\left|m_1,m_2\right\rangle$, which allows the two spins to be locally entangled. The full reference state remains a product over different dimers, but each factor is now a general two-spin quantum state. The interlayer interaction remains an explicit term in the Hamiltonian $\mathcal{H}_z=J_z\sum_i\mathbf{S}_{i,1}\cdot\mathbf{S}_{i,2}$. It is therefore not absorbed into the definition of the dimer or fixed in advance. Instead, $J_z$ determines the internal energy levels and wave functions of each dimer. For two $S=1$ moments, the isolated dimer contains total-spin sectors $T=0$, $1$, and $2$, with
$E_T=\frac{J_z}{2}\left[T(T+1)-2S(S+1)\right]$. The in-plane exchanges couple neighboring dimers and give the local excitations their momentum dependence. Varying $J_z$ therefore changes both the characteristic excitation energies and the spectral-weight distribution, even after the two outer-layer spins are grouped into a dimer. We performed the calculations using the \textsc{Sunny} \texttt{EntangledSystem} formalism, combining the two outer-layer spins into a single entangled unit, minimizing the energy of the resulting dimer array, and evaluating the trace of the dynamical spin structure factor about the optimized state. Before comparison with the experimental RIXS maps, we broadened the calculated mode intensities with a 30~meV FWHM Gaussian.

\vspace{-3mm}
\subsection{Commensurate supercell approximation}\vspace{-3mm}

We performed the entangled-dimer calculations using finite $N\times N$ real-space supercells. Periodic boundary conditions require $\left|\psi_{\mathbf{R}+N\mathbf{a}}\right\rangle=\left|\psi_{\mathbf{R}}\right\rangle$, and likewise along the second in-plane lattice direction, restricting the allowed propagation vectors to $(h,k)=(m/N,n/N)$ with integers $m$ and $n$. An arbitrary incommensurate wave vector therefore cannot be represented exactly in a finite periodic supercell. This restriction arises from the present supercell implementation rather than from the dimer description itself. Although a generalized entangled spiral could in principle describe an incommensurate state, the \texttt{EntangledSystem} calculation represents the spatial modulation explicitly within the real-space supercell. The experimental SDW vector $\delta\simeq0.31$ must therefore be approximated by a nearby rational fraction. Near the ordering vector imposed by the supercell, the calculated spectrum contains very low-energy modes and the linear spin-wave calculation becomes numerically ill-conditioned. At some nearby momenta, no stable physical spectrum is obtained. We omit these points from the maps and the resulting blank regions therefore reflect numerical limitations of the commensurate supercell calculation.

\vspace{-3mm}
\subsection{$3\times3$ dimer calculation}\vspace{-3mm}
We first considered a $3\times3$ dimer supercell, the smallest square supercell with a diagonal modulation close to the experimental SDW vector, approximated as $\delta_{\mathrm{dimer}}=1/3\simeq0.333$. Its small size also allows repeated calculations over a broad range of exchange parameters, making an approximate least-squares analysis computationally feasible. For a diagonal propagation vector $(1/3,1/3)$, the in-plane exchanges satisfy approximately $J_1\simeq J_2+2J_3$.
\begin{figure*}[h]
    \centering
    \includegraphics[width=0.9\textwidth]
    {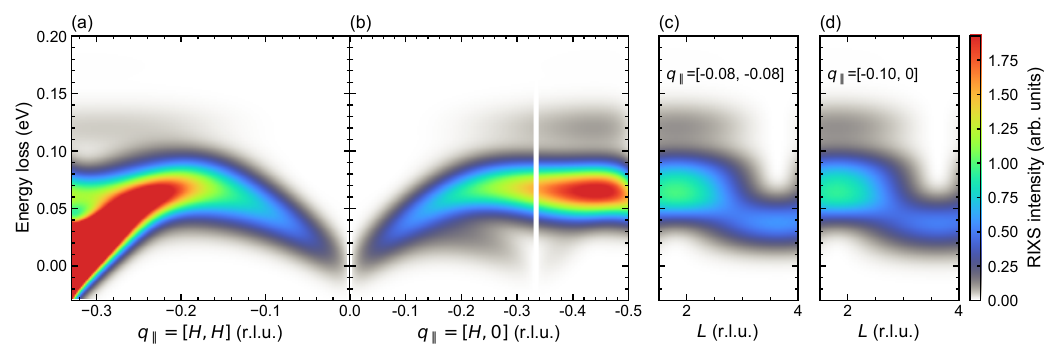}
    \caption{
    Magnetic response calculated using the $3\times3$ entangled-dimer model with $J_1=19.1$~meV, $J_2=2.5$~meV, $J_3=8.2$~meV, and $J_z=14.8$~meV.
    (a) In-plane response along $q_{\parallel}=[H,H]$.
    (b) In-plane response along $q_{\parallel}=[H,0]$.
    (c) $L$ dependence at fixed $q_{\parallel}=(-0.08,-0.08)$.
    (d) $L$ dependence at fixed $q_{\parallel}=(-0.10,0)$.
    The mode intensities were broadened by a Gaussian of 30~meV FWHM. The narrow vertical blank region in panel (b) marks a momentum at which the linear spin-wave calculation becomes unstable near a commensurate wave vector of the $3\times3$ supercell. The nominal $[H,H]$ response in panel (a) includes equal-weight averaging over the two structural twins described in Sec.~\ref{sec:lswt_domains}.
    }
    \label{fig:S21_dimer_3x3}
\end{figure*}
We evaluated the calculated spectra at the measured $\mathbf{q}=(H,K,L)$ points for both in-plane and both $L$-dependent momentum cuts, and linearly scaled the calculated intensity to the experiment before comparison. An approximate simultaneous least-squares optimization yielded 

\begin{equation}
J_1=19.1~\mathrm{meV}\quad J_2=2.5~\mathrm{meV}\quad J_3=8.2~\mathrm{meV}\quad J_z=14.8~\mathrm{meV}.
\end{equation}

A subsequent local refinement, which shifts the fitted parameters by less than $0.3$~meV, produces visually indistinguishable calculated spectra. Because some nearby parameter sets yield numerically unstable spin-wave solutions in the finite $3\times3$ supercell, we estimate the parameter sensitivity using local profile scans rather than assigning formal $\chi^2$ uncertainties. For $J_z$, the profile scan gives a characteristic range of approximately $14.1$--$15.6$~meV. The corresponding response is shown in Fig.~\ref{fig:S21_dimer_3x3}. The model reproduces the overall in-plane magnetic response and the main qualitative variation of spectral weight with $L$, while yielding an interlayer interaction comparable to that obtained from conventional LSWT. The main limitation of the $3\times3$ calculation is the mismatch between the imposed wave vector $1/3$ and the experimental value $\delta\simeq0.31$. Because the strongest experimental response lies near this region, the mismatch can bias the extracted exchange parameters. In Fig.~\ref{fig:S21_dimer_3x3}(b), a narrow vertical instability near $H=-1/3$ along $[H,0]$ reflects the commensurate periodicity of the $3\times3$ supercell.

\subsection{$13\times13$ dimer calculation}

\begin{figure*}[h]
    \centering
    \includegraphics[width=0.7\textwidth]
    {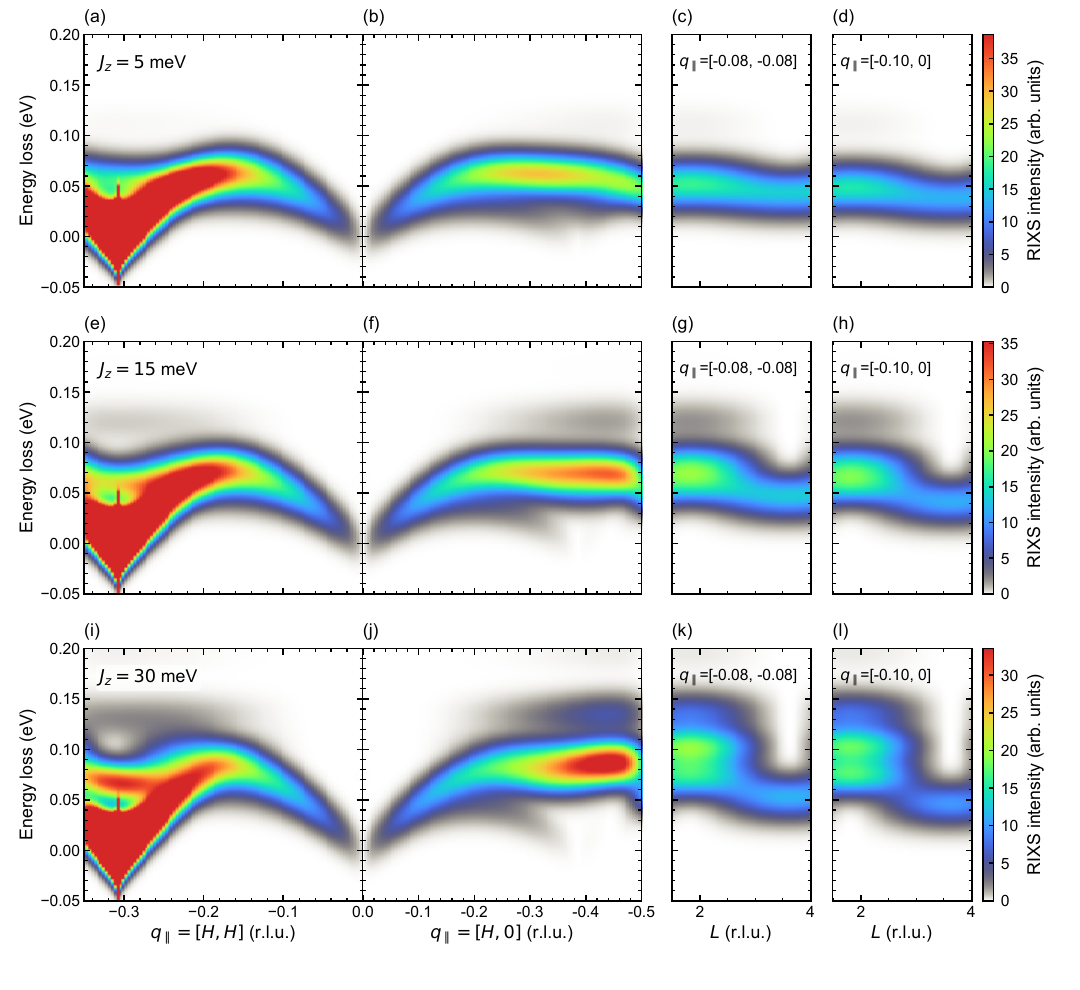}
    \caption{Magnetic response calculated using the $13\times13$ entangled-dimer model with propagation vector $\delta_{\mathrm{dimer}}=4/13$, $J_1=15$~meV, $J_2=0$, and $J_3=10.58$~meV. The rows correspond to $J_z=5$~meV in panels (a)--(d), $J_z=15$~meV in panels (e)--(h), and $J_z=30$~meV in panels (i)--(l). Within each row, the panels show the response along $q_{\parallel}=[H,H]$, along $q_{\parallel}=[H,0]$, as a function of $L$ at fixed $q_{\parallel}=(-0.08,-0.08)$ r.l.u., and as a function of $L$ at fixed $q_{\parallel}=(-0.10,0)$ r.l.u., respectively. The mode intensities were broadened by a Gaussian of 30~meV FWHM. Each row is displayed using an independently determined color scale, so absolute intensities should not be compared between different values of $J_z$. The $[H,H]$ responses in panels (a), (e), and (i) include equal-weight averaging over the two structural twins.}\vspace{-3mm}
    \label{fig:S22_dimer_13x13}
\end{figure*}
To better approximate the experimental propagation vector, we next considered a $13\times13$ dimer supercell, which permits the commensurate value $\delta_{\mathrm{dimer}}=4/13\simeq0.3077$, very close to the experimental $\delta\simeq0.31$. A substantially larger denominator would only modestly improve the wave-vector approximation while increasing the computational cost considerably. For the $13\times13$ calculation, we fixed $J_1=15~\mathrm{meV}$ and $J_2=0$, and determined $J_3$ from the fixed-wave-vector condition $J_3=-\frac{J_1}{4\cos\left(2\pi\,4/13\right)}-\frac{J_2}{2}$, giving $J_3=10.58~\mathrm{meV}$. We then calculated the response for $J_z=5$, $15$, and $30$~meV. The computational cost grows rapidly with supercell size because both the number of dimers and the number of spin-wave branches scale with the supercell area. The $13\times13$ calculation therefore contains many more branches at each momentum point than the $3\times3$ case. Even on a coarser momentum grid, the intensity calculations remain substantially more expensive, making a full least-squares optimization impractical. We therefore use the $13\times13$ results to assess the qualitative dependence of the magnetic response on $J_z$. Figure~\ref{fig:S22_dimer_13x13} compares the calculated maps for the three values of $J_z$. Increasing $J_z$ raises the characteristic energy of the dimer excitations and redistributes spectral weight among the available branches. For $J_z=5$~meV, most of the response remains at relatively low energy, whereas for $J_z=30$~meV a large fraction of the spectral weight shifts to higher-energy modes, particularly near $L\simeq2$. Among the values considered, $J_z=15$~meV most closely reproduces the energy and $L$ dependence of the dominant spectral weight, enhancing the higher-energy response near $L\simeq2$ while retaining lower-energy weight near $L\simeq3$--$4$. This comparison remains qualitative because the momentum sampling is coarse and the in-plane parameters were not independently optimized for each $J_z$.

\vspace{-3mm}
\subsection{Stripe-dimer calculation}\vspace{-3mm}
\label{sec:stripe_dimer}

\begin{figure*}[h]
    \centering
    \includegraphics[width=0.85\textwidth]{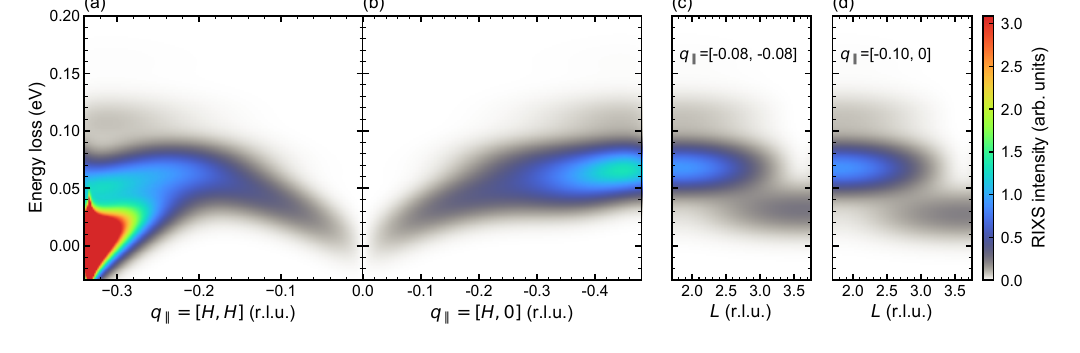}
    \caption{
    Magnetic response calculated using the stripe-dimer model with $J_1=17.9$~meV, $J_{2a}=J_{2b}=0$, $J_3=4.5$~meV, and $J_z=21.4$~meV. (a) In-plane response along $q_{\parallel}=[H,H]$. (b) In-plane response along $q_{\parallel}=[H,0]$. (c) $L$ dependence at fixed $q_{\parallel}=(-0.08,-0.08)$ r.l.u. (d) $L$ dependence at fixed $q_{\parallel}=(-0.10,0)$ r.l.u. The calculated mode intensities were broadened by a Gaussian of 30~meV FWHM.
    }
    \label{fig:S23_stripe_dimer}
\end{figure*}

As an additional comparison, we considered the stripe-dimer model used to describe magnetic excitations in trilayer Ruddlesden--Popper nickelates in Ref.~\cite{he2026layer}. The outer NiO$_2$ layers contain a period-three in-plane pattern of two magnetic $S=1$ sites followed by one spinless site, with equal contributions from two orthogonal structural twins. The exchange network includes in-plane couplings $J_1$, $J_{2a}$, $J_{2b}$, and $J_3$, together with the effective outer-layer coupling $J_z$. Following Ref.~\cite{he2026layer}, we fixed $J_{2a}=J_{2b}=0$. A simultaneous fit to the two in-plane and two $L$-dependent RIXS maps gives $J_1=17.9~\mathrm{meV}$, $J_3=4.5~\mathrm{meV}$, and $J_z=21.4~\mathrm{meV}$. The corresponding response is shown in Fig.~\ref{fig:S23_stripe_dimer}. The model reproduces the overall energy scale of the in-plane magnetic dispersion and its characteristic $L$-dependent spectral-weight variation, providing a description of the measured response based on an explicit stripe configuration.

\subsection{Comparison with the conventional LSWT description}
\label{sec:dimer_comparison}

The different spin-wave descriptions give a consistent picture of the magnetic interactions in Pr$_4$Ni$_3$O$_{10}$. The conventional LSWT fit gives $J_z=15.8$~meV, the $3\times3$ entangled-dimer fit $J_z=14.8$~meV, and the $13\times13$ calculation shows the best qualitative agreement with the measured $L$ dependence for $J_z\simeq15$~meV. The stripe-dimer model yields a somewhat larger value, $J_z=21.4$~meV. Despite differences in magnetic structure and treatment of the outer-layer spins, all models require a substantial interlayer interaction on the same energy scale as the dominant in-plane exchanges. This interaction hierarchy differs from the strongly dimerized limit reported for bilayer La$_3$Ni$_2$O$_7$, where the interlayer coupling dominates the magnetic response. In Pr$_4$Ni$_3$O$_{10}$, the results instead support a more balanced network of significant in-plane and interlayer interactions. The agreement between the conventional and entangled-dimer calculations further shows that the inferred interlayer energy scale is not strongly modified by explicitly including local quantum entanglement between the two outer-layer moments. Among the models considered, conventional LSWT provides the most direct and computationally efficient quantitative description. It represents the experimental incommensurate SDW wave vector explicitly, avoids the large real-space supercells required by the entangled-dimer calculations, and allows systematic simultaneous fitting of all four measured momentum cuts over a broad parameter range. We therefore use conventional LSWT as the primary description of the magnetic excitations, with the entangled-dimer and stripe-dimer calculations providing complementary tests of the robustness of the extracted interaction hierarchy.

\bibliography{ref} 